Influence of Biomass Emissions upon Habitability,
Biosignatures and Detectability in Earth-like Atmospheres

Stefanie Gebauer[1,2], Iva Vilović[3], John Lee Grenfell[2],
Fabian Wunderlich[3], Franz Schreier[4] and Heike Rauer[2,3,5]

**Abstract**

We investigate atmospheric responses of modeled hypothetical Earth-like planets in the habitable zone of the M-dwarf AD Leonis to reduced oxygen ($O_2$), removed biomass ("dead" Earth), varying carbon dioxide ($CO_2$) and surface relative humidity (sRH). Results suggest large $O_2$ differences between the reduced $O_2$ and "dead" scenarios in the lower but not the upper atmosphere. Ozone ($O_3$) and nitrous oxide ($N_2O$) also show this behavior. Methane depends on hydroxyl (OH), its main sink. Abiotic production of $N_2O$ occurs in the upper layers. Chloromethane ($CH_3Cl$) decreases everywhere on decreasing biomass. Changing $CO_2$ (from x1 to x100 present atmospheric level (PAL)) and surface relative humidity (sRH) (from 0.1% to 100%) does not influence $CH_3Cl$ as much as lowering biomass. Therefore, $CH_3Cl$ can be considered a good biosignature. Changing sRH and $CO_2$ has a greater influence on temperature than $O_2$ and biomass alone. Changing the biomass produces ~6 kilometer (km) in effective height (H) in transmission compared with changing $CO_2$ and sRH (~25km). In transmission $O_2$ is discernible at 0.76 µm for >0.1 PAL. The $O_3$ 9.6 µm band was weak for the low $O_2$ runs and difficult to discern from "dead" Earth", however $O_3$ at 0.3 µm could serve as an indicator to distinguish between reduced $O_2$ and "dead" Earth. Spectral features of $N_2O$ and $CH_3Cl$ corresponded to some km H. $CH_4$ could be detectable tens of

1       Corresponding author stefanie.gebauer@dlr.de

2       Institute for Planetary Research (PF), German Aerospace Centre (DLR), Rutherfordstr. 2, 12489 Berlin, Germany

3       Centre for Astronomy and Astrophysics (ZAA), Berlin Institute of Technology (TUB), Hardenbergstr. 36, 10623 Berlin, Germany

4       Remote Sensing Technology Institute (IMF), German Aerospace Centre (DLR), 82234 Oberpfaffenhofen, Germany

5       Institute for Geological Sciences, Planetology and Remote Sensing, Free University of Berlin (FUB), Malteserstr. 74-100, 12249 Berlin, Germany parsecs away with ELT except for the $10^{-4}$ and $10^{-6}$ PAL $O_2$ scenarios. $O_2$ is barely detectable for the 1 PAL $O_2$ case and unfeasible at lower abundances.

# 1. Introduction

## 1.1. Motivation

The end of the last century brought with it the technological possibility of detecting planets outside our Solar System. The first extrasolar planet orbiting a main-sequence star was discovered by Mayor & Queloz (1995). Up until today, more than 4000 extrasolar planets have been confirmed (www.exoplanet.eu). An increasing number of these planets fall in the range of so-called super-Earths orbiting M-dwarf stars (e.g., Affer et al., 2016; Gillon et al., 2016, 2017, Anglada-Escudé et al., 2016). Their discovery is just the beginning; what are the characteristics of these worlds? Are any signs of extraterrestrial life present? Is Earth a unique planet or is life a common occurrence in the Universe? Applying various detection methods, astronomers have been able to determine exoplanetary properties (such as orbital period, mass and radius). Exoplanetary atmospheres could be very diverse (e.g. Leconte et al. 2015; Grenfell et al., 2020). Such things have motivated discussions regarding the conditions to be met by a planet in order to enable and accommodate the development of life (Davies, 1995).

Due to the lack of knowledge of other possible life forms, such discussions centered mostly on life as we know it from our own planet. On Earth, molecular oxygen ($O_2$) is central for the formation of ozone ($O_3$) and is mostly produced by oxygenic photosynthesis, nitrous oxide ($N_2O$) from bacteria and fungi in soils and oceans and methyl chloride ($CH_3Cl$) which comes from seaweed and other vegetation. These molecules can be considered as biosignatures because their abundances on Earth require a biological agent. Methane ($CH_4$) is mainly produced by methanogenic bacteria but is not considered to be a biosignature because it is also produced abiotically via volcanic and metamorphic outgassing. It is sometimes referred to as a "bio-indicator" (i.e. a possible hint for life, but more information is required). However, an individual gas alone (present in Earth's atmosphere) is likely insufficient as a confirmation of a biosignature (e.g. Schwieterman et al., 2018). For example, the simultaneous presence of $O_3$, water ($H_2O$) and carbon dioxide ($CO_2$) in the atmosphere, also called the "triple-signature" (Selsis et al., 2002), is considered a good indication of life. The simultaneous detection of large amounts of $CH_4$, which is a reducer, and $O_2$, which is an oxidizer, would indicate a continuous production via biology (Sagan et al., 1993). Furthermore, complex life as we know it is not possible without $O_2$, which is produced mostly by biology on Earth (Catling et al., 2005; McKay, 2014). Some works (see e.g. Segura et al., 2007; Tian et al., 2014; Wordsworth & Pierrehumbert, 2014; Domagal-Goldman et al., 2014; Luger & Barnes, 2015; see also Harman et al. 2018) proposed abiotic build-up of $O_2$ in certain exoplanetary atmospheres, leading to "false-positive" (i.e. false confirmations of life) detections in the spectrum (see e.g. Selsis et al., 2002, Schwieterman et al., 2018). If life is overlooked in detection, this is considered a "false-negative" (see e.g. Reinhard et al., 2017). It is therefore necessary to understand the surrounding environmental context in order to draw conclusions about the biosignature-status of an atmospheric chemical species.

In order to better understand the impact of life upon a planet's atmosphere, it is informative to calculate a so-called "dead" Earth scenario as a benchmark for comparison. Central studies that could be considered the starting point of the modern-day examination and comparison of a hypothetical "dead" Earth to the present day Earth are Lovelock & Margulis (1974) and Margulis & Lovelock (1974) who explore the concept that a planet's biosphere actively stabilizes its own presence via complex biogeochemical feedbacks, the so-called "Gaia hypothesis". Considering the atmosphere to be a circulatory system of the biosphere, they discuss approaches for comparing living with non-living atmospheres using Earth as an example. Firstly they "delete" life from the present Earth and consider the response in the atmospheric composition and climate. Secondly they consider the evolution of the Earth assuming that life had never evolved.  The first approach is rather less challenging to realize yet it possibly overlooks subtle time-dependent feedbacks of the biosphere upon the atmosphere. These earlier studies claimed that if life were "deleted" from present Earth, then nearly all $O_2$ and molecular nitrogen ($N_2$) would disappear due to, among other things, sinks due to e.g. lightning and high energy radiation from the Sun (see also Yung & DeMore, 1999, and references therein). This would leave nitrogen in its chemically stable form: the hydrated nitrate ion in the oceans (see however studies by e.g. Kasting, 1993; Ranjan et al., 2019, who argue that $N_2(g)$ would be re-gassed into the atmosphere). $CO_2$ would become the dominant constituent of the hypothetical atmosphere ranging anywhere between 0.3 millibar (mbar) and 1000 mbar, as its removal rate by geochemical processes would slow down without life due to less weathering and 'carbon-rain' in the oceans (Yung & DeMore, 1999). Thirdly, they investigated a scenario where life never developed; in this case Earth's atmosphere, similarly to Mars and Venus, would have begun in reduced form, rich in molecular hydrogen ($H_2$), $CH_4$ and ammonia ($NH_3$). The primordial atmosphere would have escaped to space and the atmosphere would then have drifted towards a less reduced state, rich in carbonates, nitrates and sulfates. Morrison & Owen (2003) suggested that the whole $CO_2$ budget of modern Earth (approximately 69 bar) could be returned to the atmosphere in the case of an abiotic, i.e. "dead" Earth. The study by Franck et al. (2006) applies a minimal model for Earth's global carbon cycle in order to derive the ultimate life span of the biosphere defined by the extinction of procaryotes in about 1.6 Gigayear (Gyr). This result is also supported by the approach of de Sousa Mello & Friaca (2020), in which they analyze how long it would take for the biosphere to be extinguished in the future e.g. due to the increase in solar luminosity. Their results indicate that the biosphere would collapse due to high temperatures in approximately 1.63 (+ 0.14, -0.05) Gyr even before the Sun becomes a red giant.

The last few years have seen an increasing focus upon numerical model calculations investigating the climate and chemical composition of Earth-like atmospheres. Various models are used to simulate atmospheric chemical compositions, as well as a planet's potential habitability. The model output can be used to calculate hypothetical spectra, which would aid in interpreting observed transit spectroscopy. Some 1-dimensional (1D) modeling studies which focus on the observations of Earth for the interpretation of biosignatures include e. g. Arnold et al. (2002), Seager et al. (2005), Sagan et al. (1993), Hurley et al. (2014), Schwieterman et al. (2015). Various climate and/or photochemical model studies have also been used to simulate how the stellar spectrum influences the atmospheric composition of hypothetical planets, including e.g. Kasting et al. (1993), Kasting (1997), Selsis et al. (2002), Pavlov & Kasting (2002), Domagal-Goldman et al. (2011), Segura et al. (2003, 2005, 2007, 2010), Rauer et al. (2011), Arney et al. (2016), Yung et al. (1988), Gao et al. (2015), Hu et al. (2012), Hu & Seager (2014), Grenfell et al. (2014), Gebauer et al. (2018a), Scheucher et al. (2018), Wunderlich et al. (2019, 2020) and Scheucher et al. (2020). Studies that model Earth's atmospheric composition over geological time include e.g. Meadows (2005), Kaltenegger et al. (2007), and Gebauer et al. (2017, 2018b) among others. Whereas 1D models are useful for large parameter studies in exoplanet science, 3-dimensional (3D) models are applied to investigate e.g. global transport, the hydrological cycle and clouds. A notable caveat in 3D climate calculations is that they require a large set of boundary conditions, such as continental distribution, orography, obliquity, rotation rate, and oceanic heat transport. These are, in addition to fundamental parameters such as atmospheric mass and composition, mostly not known for potentially habitable rocky extrasolar planets. Despite this, much progress has been made in 3D studies in recent years. Some central studies of potentially habitable planets and early Earth include e.g. Robinson et al. (2011), Leconte et al. (2013a, 2013b), Yang et al. (2013, 2014), Wolf & Toon (2013), Charnay et al. (2013), Kunze et al. (2014), Shields et al. (2014, 2016), Kopparapu et al. (2016), Way et al. (2016), Wolf et al. (2017) and more. Not only have atmospheric and spectral models increasingly become a powerful tool for investigating atmospheric properties, but their ability to model the atmospheres of terrestrial extrasolar planets has been steadily improved, making them indispensable for future research on planetary habitability.

Previous major modeling efforts which calculated synthetic spectra of hypothetical Earth-like extrasolar planets were performed by e.g. Selsis (2000); Selsis et al. (2002); Des Marais et al. (2002); Segura et al. (2003, 2005); Tinetti (2006); Ehrenreich et al. (2006); Kaltenegger et al. (2007); Kaltenegger & Traub (2009); Kaltenegger & Sasselov (2010); Rauer et al. (2011); von Paris et al. (2011), Hedelt et al. (2013), Snellen et al. (2013), Rodler & López-Morales (2014), Bétrémieux & Kaltenegger (2014), Misra & Meadows (2014), Stevenson et al. (2016), Barstow et al. (2016), Barstow & Irwin (2016), Rugheimer & Kaltenegger (2018), Kaltenegger et al. (2019), Lin & Kaltenegger (2019), Schreier et al., (2020) and Fauchez et al. (2020). Studies of the Earth that explore the sensitivity of its spectra to the temperature structure (a hot and cold scenario) and different evolutionary stages have also been published, e.g. Schindler & Kasting (2000); Selsis (2000); Segura et al. (2003); Kaltenegger et al. (2007); Meadows (2005), Kaltenegger et al. (2019). Checlair et al. investigate potential future detections of $O_2$ and $O_3$ on Earth-like planets.

Previous studies with state-of-the-art models have generally lacked comparisons with "dead" Earth scenarios as a benchmark for biosignature assessment and it would be beneficial in future for similar scenarios to be performed with other models in the literature.

## 1.2. Contribution of this Work

The oldest known fossils suggest that organisms were present 3.5 billion years ago (Noffke et al., 2013), indicating that life had been present for at least a billion years before the Great Oxidation Event (GOE) at 2.4-2.3 Gyr. This suggests it is of importance to explore potential signals from hypothetical Earth-like planets with $O_2$ ranging from the pre-GOE atmospheric levels up to the modern Earth atmospheric levels. "Earth-like" in our study refers to rocky worlds with internal structure, instellation and bulk atmosphere properties broadly comparable to Earth. Furthermore, studying hypothetical Earth-like worlds in which life is not present (i.e. "dead Earths") can serve as an important benchmark for comparison when interpreting biosignature candidates, and can shed light on how life could modify a planet's atmospheric composition and climate. Varying additional parameters, such as atmospheric $CO_2$ content and surface relative humidity (sRH), can also be a useful tool in interpreting an (exo)planetary atmosphere's response since they are key gases affecting habitability and could potentially mask absorption features of other species in a planet's atmospheric spectrum. They can have a large impact on a planet's climate, for which the contribution from life is generally not well constrained. Therefore, this work aims to explore the atmospheric and spectral response of a hypothetical Earth-like planet to changing parameters such as the central star (in this case AD Leonis), lowering the $O_2$ content (both as part of an $O_2$ parameter study as well as in combination with varying other parameters), changing the sRH and $CO_2$ content, as well as removing biomass emissions (i.e. modeling an "abiotic" or "dead" Earth). These large differences and uncertainties are associated with the complicated way in which life influences the Earth system.

Our studied planet orbits in the habitable zone (HZ) around a cool M-dwarf star. These objects are favored targets when searching for life outside the Solar System although some of their characteristics could disfavor the development and evolution of life (see e.g. Shields et al., 2016; Gebauer et al, 2018a and references therein). We perform scenarios for Earth-like planets orbiting in the HZ at 0.153 Astronomical unit (AU) around the well-studied M-dwarf star AD Leonis (AD Leo) where the total stellar wavelength-integrated energy input equals that of 1 solar constant (1367 W m$^{-2}$). We apply the Coupled Atmosphere Biogeochemical (CAB) model (Gebauer et al. 2017, 2018a, 2018b) and the "Generic Atmospheric Radiation Line-by-line Infrared-microwave Code" (GARLIC) (Schreier et al., 2014; Schreier et al., 2018a, 2018b). Furthermore, we investigate to what extent calculated spectral signals for atmospheric biosignatures could be attributed to biological activity as opposed to abiotic atmospheric production and explore the uncertainties associated with varying sRH, $O_2$ abundance and greenhouse gases. We also estimate detectability and S/N of some key spectral features assuming the instrumental setup proposed for the Extremely Large Telescope (ELT). This is the first such study to our knowledge exploring the responses of modern numerical atmospheric models to the above scenarios.

The paper is structured as follows: Section 2 describes briefly the models used. Section 3 gives an overview of the simulated Earth-like planet scenarios around

AD Leo, whereas Section 4 presents and discusses the results. Our main conclusions are presented in Section 5.

## 2. Model Description

### 2.1. Atmospheric Column Model

In this study we use the 1D global mean cloud-free, steady state atmospheric module of the Coupled Atmospheric Biogeochemical (CAB) model in order to investigate the atmospheric responses upon removing all biomass emissions. The original atmospheric module is based on the work by e.g. Kasting et al. (1984), Segura et al. (2003), Grenfell et al. (2007a, 2007b), Rauer et al. (2011), Grenfell et al. (2011), von Paris et al. (2015) and was further updated by Gebauer et al. (2017, 2018a, 2018b). The model is composed of climate and photochemistry modules calculating temperature, water, and chemical species profiles as a function of boundary conditions such as the stellar input spectrum, biogenic surface fluxes and deposition velocities. The atmosphere in the climate module is divided into 52 adaptive pressure levels whereas the chemistry module uses 64 equidistant altitude levels. In the climate module the radiative transfer is separated into a short wavelength region ranging from 237.6 nanometers (nm) to 4.545 micrometers (µm) with 38 wavelength bands for incoming stellar radiation and a long wavelength region from 1 to 500 µm in 25 bands for planetary and atmospheric thermal radiation (von Paris et al. 2015). Rayleigh scattering by $N_2$, $O_2$, $H_2O$, $CO_2$, $CH_4$, $H_2$, He and CO is used by applying the two-stream radiative transfer method based on Toon et al. (1989). Molecular absorption in the short wavelength range by the major absorbers $H_2O$, $CO_2$, $CH_4$ and $O_3$ is considered. Molecular absorption of thermal radiation by $H_2O$, $CO_2$, $CH_4$ and $O_3$ and continuum absorption by $N_2$, $H_2O$ and $CO_2$ are included (von Paris et al. 2015). The water vapor concentrations in the troposphere are calculated using the relative humidity profile of the Earth taken from Manabe & Wetherald (1967). The chemistry module includes 55 species with more than 200 chemical reactions. It calculates the profiles of $CO_2$, $O_2$ and $N_2$ through the photochemical network instead of fixing the profiles to isoprofiles as done in earlier works, e.g. by Segura et al. (2003, 2005), Rauer et al. (2011), etc. The photolysis rates are calculated in the wavelength range of 121.4 to 855 nm. For the effective $O_2$ cross sections in the Schumann-Runge bands we use the values from Murtagh (1988) as described in Gebauer et al. (2018b). In order to reproduce the modern Earth ozone column of ~300 Dobson Units (DU) a mean solar zenith angle of 53° in the photochemistry module is applied. We set the surface albedo to a value of 0.21 in order to reproduce the modern Earth surface temperature of 288.15 K. For a more detailed description of the CAB model we refer to Gebauer et al. (2017, 2018a, 2018b) and references therein.

## 2.2. Radiative Transfer Model

The "Generic Atmospheric Radiation Line-by-line Infrared-microwave Code" (GARLIC) described in Schreier et al. (2014), Schreier et al. (2018a, 2018b) takes pressure, temperature and composition output by the CAB Model in order to calculate synthetic transmission spectra. High resolution infrared (IR) and microwave (MW) radiative transfer calculations by means of "line-by-line" models are commonly applied in Earth and (exo)planetary science. GARLIC has been developed for arbitrary observational geometries, as well as instrumental fields-of-view and spectral response functions and is based on MIRART-SQuIRRL (Schreier & Schimpf, 2001; Schreier & Böttger, 2003), which has been used for radiative transfer modeling in previous studies, e.g. Rauer et al. (2011) and Hedelt et al. (2013). The line parameters over the whole wavelength range are taken from the HITRAN 2016 database (Gordon et al. 2017). In addition, the Clough-Kneizys-Davies continuum model (CKD; Clough et al. 1989) and Rayleigh scattering are considered (Murphy 1977; Clough et al. 1989; Sneep & Ubachs 2005; Marcq et al. 2011). $O_3$ cross sections below 1100 nm are implemented from Serdyuchenko et al. (2014). Note that not all of the species in CAB are considered to be spectroscopically relevant for Earth-like atmospheres (see e.g. Schreier et al. 2018), i.e. GARLIC uses the 23 chemical species (which HITRAN 2016 and CAB have in common) as input. Transmission spectra are calculated for different heights corresponding to a limb geometry and through 64 atmospheric layers and are summed up according to Schreier et al (2018a) to derive effective height spectra.

## 2.3. Spectral Detectability Method

We assume an Earth-like planet placed in the HZ of AD Leo such that it receives an instellation equal to that of the modern Earth and then calculate the number of transits required for a signal-to-noise (S/N) equal to 5.0, hereafter "$S/N_5$". Spectral signals for atmospheric species are taken from section 4.2 whereas photon and instrument noise are calculated assuming the ELT High Resolution Spectrograph (HIRES) instrument:

(a) For $CH_4$ calculated up to 50 parsec (pc).

(b) For $O_2$ calculated up to 25 pc.

More details of the approach can be found in Wunderlich et al. (2020).

To calculate the *S/N* of a planetary atmospheric feature, *S/N*atm for a single transit, we first calculate the *S/N* of the star, *S/N*s , integrated over one transit and then multiply this value with the atmospheric signal, *S*atm:

$$S/N_{atm} = S/N_s \cdot S_{atm} / \sqrt{2} \qquad (1)$$

The factor $1/\sqrt{2}$ in equation (1) accounts for the fact that the star is observed for both in-transit and out-of-transit. The transit duration for Earth around AD Leo is 3.02 hours (see e. g. Wunderlich et al. 2019). *S*atm is the "relative" transit depth, i.e. the wavelength dependent transit depth minus minimum transit depth of the considered wavelength range. The atmospheric contribution of the transit depth = $(R_{Earth} + h_e)^2/R_{adleo}^2 - (R_{Earth}^2/R_{adleo}^2)$ and $R_{adleo} = 0.39\ R_{sun}$ (Reiners et al., 2009), where $h_e$ is the effective height, calculated with the GARLIC line-by-line model (refer to equations (5), (6), (7) and (8) in Wunderlich et al., 2020) and $R_{Earth}$, $R_{adleo}$ and $R_{sun}$ are the radii of Earth, AD Leo and the Sun, respectively.

The *S/N*s of AD Leo obtained with the ELT instrument is calculated with the ESO Exposure Time Calculator (ETC) as described in Liske (2008). The AD Leo spectrum is taken from Segura et al. (2003) and scaled to the K-band magnitude of 4.593 (Cutri et al. 2003) to obtain the input flux distribution. We assume a telescope with a diameter of 39 m, located at Cerro Armazones and an average throughput of 10%. The sky conditions are set to a constant airmass of 1.5 and a precipitable water vapor (PWV) of 2.5 (Liske, 2008). The number of transits for $O_2$ (1.24 – 1.3 µm) and $CH_4$ (2.1 – 2.5 µm) are calculated assuming ELT HIRES with R=100,000 (Marconi et al., 2016) using the cross-correlation technique (see e.g. Snellen et al., 2013; Rodler & López-Morales et al., 2014; Birkby et al., 2013; Brogi et al., 2018; López-Morales et al., 2019). We calculate the number of transits, necessary to reach an *S/N* of 5, assuming that all transits improve the *S/N*s perfectly (see discussion on noise sources in Wunderlich et al., 2020). The *S/N* is calculated from equation (9) in Wunderlich et al. (2020).

## 3. Scenarios

### 3.1. AD Leo Scenarios

In this work we simulate Earth-like planets with 1 Earth radius and mass placed in the HZ (at 0.153 AU) around the M-dwarf star AD Leo so that they receive the same instellation of 1367 W m⁻² as modern Earth from the Sun. Tables 1 and 2 summarize the boundary conditions for the AD Leo reference (control) run whereas Tables 3 and 4 summarize the different model scenarios which are separated into two "families" of runs. These include (1) a set of low oxygen (i.e. $10^{-6}$ Present Atmospheric Level (PAL) $O_2$ fixed at the surface in order to account for a weak photosynthetic biosphere; Table 3) and (2) "dead" Earth (Table 4) runs. In the case of the low oxygen runs, boundary conditions of species such as $N_2$, $CH_4$, $N_2O$, $H_2$, CO and $CH_3Cl$ have not been changed relative to the control run in order to simulate the presence of a biosphere. For the "dead" Earth scenarios the surface fluxes of $CH_4$, $N_2O$, $H_2$ and $CH_3Cl$ have been strongly lowered to simulate the deletion of methanogenic microbes, the absence of (de)nitrifying bacteria, the halted consumption by enzymes and the deletion of associated vegetation, respectively. Furthermore, based on Tian et al. (2014) the deposition velocities of CO and $O_2$ have both been lowered to $10^{-6}$ cm s$^{-1}$.

| Gas | AD Leo control | | |
|---|---|---|---|
| | Biological Surface flux [molec./(cm²s)] | Deposition velocity [cm/s] | Volume mixing ratio [vmr] |
| $N_2$ | | | 78% |
| $O_2$ | | | 21% |
| $H_2O$ | | | 78% sRH |
| $CO_2$ | | | 355 ppm |
| $CH_4$ | $8.6 \cdot 10^{10}$ | | |
| $N_2O$ | $1.3 \cdot 10^{9}$ | | |
| $H_2$ | | $2.5 \cdot 10^{-3}$ | |
| $CO$ | $2.0 \cdot 10^{11}$ | | |
| $CH_3Cl$ | $2.0 \cdot 10^{8}$ | | |

**Table 1:** *Boundary conditions – biological surface flux, deposition velocity, constant surface volume mixing ratio – shown for the relevant chemical species of the AD Leo control run simulated in this study. sRH stands for surface relative humidity. Greenhouse gas concentrations (e.g. $CO_2$ = 355 parts per million (ppm)) unless indicated otherwise correspond to 1990 Earth conditions to compare with numerous similar studies in the literature (e.g. Segura et al., 2003; Rauer et al., 2011).*

| Gas | Volcanic flux [molec./cm²s] | Metamorphic flux [molec./cm²s] | Total geological flux [molec./cm²s] | Reference |
|---|---|---|---|---|
| $H_2S$ | $1.5 \cdot 10^{8}$ | | $1.5 \cdot 10^{8}$ | Halmer (2002) |
| $SO_2$ | $8.7 \cdot 10^{8}$ | | $8.7 \cdot 10^{8}$ | Halmer (2002) |
| $CO_2$ | $3.1 \cdot 10^{10}$ | | $3.1 \cdot 10^{10}$ | Mörner & Ethiope (2002) |
| $CH_4$ | $9.3 \cdot 10^{8}$ | $1.48 \cdot 10^{10}$ | $1.5 \cdot 10^{10}$ | Kvenvolden & Rogers (2005) Claire et al. (2006) |
| $H_2$ | $1.8 \cdot 10^{10}$ | $7.2 \cdot 10^{10}$ | $9.0 \cdot 10^{10}$ | Holland (2002) Claire et al. (2006) |
| $CO$ | $2.0 \cdot 10^{8}$ | | $2.0 \cdot 10^{8}$ | Zahnle et al. (2006) |

**Table 2**: *Geological (volcanic + metamorphic) fluxes considered in the photochemical calculations.*

| Gas | Low Oxygen | | | Low Oxygen High $CO_2$ | | |
|---|---|---|---|---|---|---|
| | Biological Surface flux [molec./(cm²s)] | Deposition velocity [cm/s] | Volume mixing ratio [vmr] | Biological Surface flux [molec./(cm²s)] | Deposition velocity [cm/s] | Volume mixing ratio [vmr] |
| $N_2$ | | | fill gas | | | fill gas |
| $O_2$ | | | $10^{-6}$ PAL | | | $10^{-6}$ PAL |
| $H_2O$ | | | 78% sRH | | | 78% sRH |
| $CO_2$ | | | 355 ppm | | | 100 PAL |
| $CH_4$ | $8.6 \cdot 10^{10}$ | | | $8.6 \cdot 10^{10}$ | | |
| $N_2O$ | $1.3 \cdot 10^{9}$ | | | $1.3 \cdot 10^{9}$ | | |
| $H_2$ | | $2.5 \cdot 10^{-3}$ | | | $2.5 \cdot 10^{-3}$ | |
| $CO$ | $2.0 \cdot 10^{11}$ | | | $2.0 \cdot 10^{11}$ | | |
| $CH_3Cl$ | $2.0 \cdot 10^{8}$ | | | $2.0 \cdot 10^{8}$ | | |
| Gas | Low Oxygen wet | | | Low Oxygen dry | | |

| | Biological Surface flux [molec./(cm²s)] | Deposition velocity [cm/s] | Volume mixing ratio [vmr] | Biological Surface flux [molec./(cm²s)] | Deposition velocity [cm/s] | Volume mixing ratio [vmr] |
|---|---|---|---|---|---|---|
| $N_2$ | | | fill gas | | | fill gas |
| $O_2$ | | | $10^{-6}$ PAL | | | $10^{-6}$ PAL |
| $H_2O$ | | | 100% sRH | | | 0.1% sRH |
| $CO_2$ | | | 355 ppm | | | 355 ppm |
| $CH_4$ | $8.6 \cdot 10^{10}$ | | | $8.6 \cdot 10^{10}$ | | |
| $N_2O$ | $1.3 \cdot 10^{9}$ | | | $1.3 \cdot 10^{9}$ | | |
| $H_2$ | | $2.5 \cdot 10^{-3}$ | | | $2.5 \cdot 10^{-3}$ | |
| $CO$ | $2.0 \cdot 10^{11}$ | | | $2.0 \cdot 10^{11}$ | | |
| $CH_3Cl$ | $2.0 \cdot 10^{8}$ | | | $2.0 \cdot 10^{8}$ | | |

**Table 3:** *Low oxygen model scenarios with surface biomass simulated in this study including all boundary conditions – biological surface flux, deposition velocity, constant surface volume mixing ratio (vmr) – shown for the relevant chemical species. Blue shading here in the headers (and in subsequent Figures) indicates scenarios where life is present. Low oxygen indicates that oxygen is reduced from 1 Earth Present Atmospheric Level (PAL) to $10^{-6}$ Earth PAL. Lightly-yellow shaded cells indicate which species parameters have been changed compared to the AD Leo control run. "Fill gas" indicates the fill gas assumption for $N_2$ as used in e.g. Segura et al. (2003). For the high $CO_2$ runs (upper right column block) the $CO_2$ vmr has additionally been increased to 100 PAL. For the wet and dry runs (lower 2 column blocks) the surface relative humidity is fixed to 100% and 0.1% respectively in addition to having the lowered $O_2$ vmr.*

| Gas | Low Oxygen | Low Oxygen High $CO_2$ |
|---|---|---|

| Gas | Biological Surface flux [molec./(cm²s)] | Deposition velocity [cm/s] | Volume mixing ratio [vmr] | Biological Surface flux [molec./(cm²s)] | Deposition velocity [cm/s] | Volume mixing ratio [vmr] |
|---|---|---|---|---|---|---|
| $N_2$ | | | fill gas | | | fill gas |
| $O_2$ | | $10^{-6}$ | | | $10^{-6}$ | |
| $H_2O$ | | | 78% sRH | | | 78% sRH |
| $CO_2$ | | | 355 ppm | | | 100 PAL |
| $CH_4$ | 1 | | | 1 | | |
| $N_2O$ | 1 | | | 1 | | |
| $H_2$ | | $10^{-6}$ | | | $10^{-6}$ | |
| CO | | $10^{-6}$ | | | $10^{-6}$ | |
| $CH_3Cl$ | 1 | | | 1 | | |

| Gas | Low Oxygen wet | | | Low Oxygen dry | | |
|---|---|---|---|---|---|---|
| | Biological Surface flux [molec./(cm²s)] | Deposition velocity [cm/s] | Volume mixing ratio [vmr] | Biological Surface flux [molec./(cm²s)] | Deposition velocity [cm/s] | Volume mixing ratio [vmr] |
| $N_2$ | | | fill gas | | | fill gas |
| $O_2$ | | $10^{-6}$ | | | $10^{-6}$ | |
| $H_2O$ | | | 100% sRH | | | 0.1% sRH |
| $CO_2$ | | | 355 ppm | | | 355 ppm |
| $CH_4$ | 1 | | | 1 | | |
| $N_2O$ | 1 | | | 1 | | |
| $H_2$ | | $10^{-6}$ | | | $10^{-6}$ | |
| CO | | $10^{-6}$ | | | $10^{-6}$ | |

| CH₃Cl | 1 | | | 1 | | |
|---|---|---|---|---|---|---|

**Table 4**: *As for Table 3 but for "Dead" model scenarios without biomass simulated in this study. Yellow shading here in the headers (and in subsequent Figures) indicates scenarios where life has been removed ("dead Earth").*

## 3.2. Modern Earth around the Sun

For comparison with Table 1 we performed a control run for the modern Earth orbiting the Sun at 1 AU.

## 3.3. Varying Oxygen Scenarios

For the spectral analysis (see below) additional AD Leo scenarios were performed in which surface $O_2$ was varied step-by-step. The goal here was to determine the minimum necessary atmospheric $O_2$ concentration in order for $O_2$ features to be evident in a transmission spectrum. In these additional scenarios the $O_2$ content of the AD Leo control run with 1 PAL $O_2$ in Table 1 was lowered step-by-step to 0.1 PAL, $10^{-2}$ PAL, $10^{-4}$ PAL and $10^{-6}$ PAL $O_2$. The value of 0.1 PAL $O_2$ approximately corresponds to the $O_2$ content on Earth during the Neoproterozoic Oxygenation Event (NOE) at about 800 million years ago. The remaining $O_2$ values considered correspond to post- and pre-GOE $O_2$ concentrations. Understanding the evolution of atmospheric $O_2$ on Earth is crucial for understanding atmospheric biosignatures on Earth-like exoplanets.

## 4. Results

In the following section the focus lies on presenting model responses of important atmospheric biosignatures (e.g. $O_2$, $O_3$, $N_2O$, $CH_3Cl$) and related compounds (e.g. $H_2O$, $CO_2$, $CH_4$). Chemical profiles and atmospheric temperature responses are presented in Section 4.1. Resulting synthetic transmission are presented in Section 4.2.

## 4.1. Atmospheric profiles

### 4.1.1. Molecular Oxygen ($O_2$) and Ozone – ($O_3$)

Figure 1 presents $O_2$ (left) and $O_3$ (right) vmr profiles for the low oxygen (blue lines) and "dead Earth" (yellow lines) scenarios from Table 3 and 4. For comparison the same species are plotted for the modern Earth around the Sun control (grey lines) and the AD Leo control (black lines). For $O_2$ the AD Leo control run as well as the modern Earth around the Sun both exhibit an isoprofile of 21% vmr throughout the atmosphere, which indicates $O_2$ has a long chemical lifetime and is influenced primarily by transport in both runs. The ozone layer of modern Earth around the Sun is mostly reproduced (compare black and grey lines, right panel). The solid black curve shows the AD Leo control run, which suggests a similar $O_3$ layer peaking around 0.01 bar. Earth-like planets can experience strong fluxes around active M-dwarf stars like AD Leo at wavelengths shorter than ~200 nm, resulting in stimulated $O_2$ photolysis and hence increased $O_3$ concentrations. Generally, for all other scenarios considered, the vertical behavior of $O_3$ is similar to that of $O_2$ since $O_2$ is a major source for $O_3$ via the reaction: $O_2 + O + M \rightarrow O_3 + M$. We therefore focus on describing the vertical behavior of $O_2$ in the following text. The low oxygen Earth runs show a distinct minimum between 0.1 bar and 0.01 bar (except for the high $CO_2$ run). Further investigation suggested that this minimum is related to a maximum in hydrogen oxides ($HO_x = OH + HO_2 + H$) vmr, mainly produced via $H_2O$ photolysis by Lyman-alpha radiation (see Figure 6 (b)). Low oxygen leads to more UV in the upper atmosphere which favors $H_2O$ photolysis, hence $O_2$ destruction in this pressure region occurs due to a high abundance in H which removes $O_2$ via $H + O_2 + M \rightarrow HO_2 + M$. However, for the low oxygen high $CO_2$ scenario (blue dotted line) the enhanced $CO_2$ has a UV shielding effect and the $O_2$ minimum is not present. This leads to enhanced $O_2$ vmr for this case compared to the other blue curves midway between 0.1 bar and 0.01 bar, but not in the upper atmosphere where the sink species H is enhanced due to a higher $H_2O$ abundance. The low $O_2$ dry Earth scenario with 0.1% sRH (blue dashed-dotted line) stands out from the others at lower pressures of ~0.01 bar where it features particularly low $O_2$ vmr. The low $O_2$ (hence low UV shielding) conditions favor $HO_x$ production which leads to a peak in the vmr of the $O_2$-destroying species H in that pressure region (see Figure 6 (b)). Increasing $O_2$ toward the upper layers ($p < 0.01$ bar) for the low oxygen runs is consistent with the downwards diffusion of O produced by $CO_2$ photolysis in the upper atmosphere, also shown in earlier works such as Gebauer et al. (2018a). For the low oxygen runs, in summary, the atmosphere destroys $O_2$ from the surface up, while it produces $O_2$ abiotically from the top down.

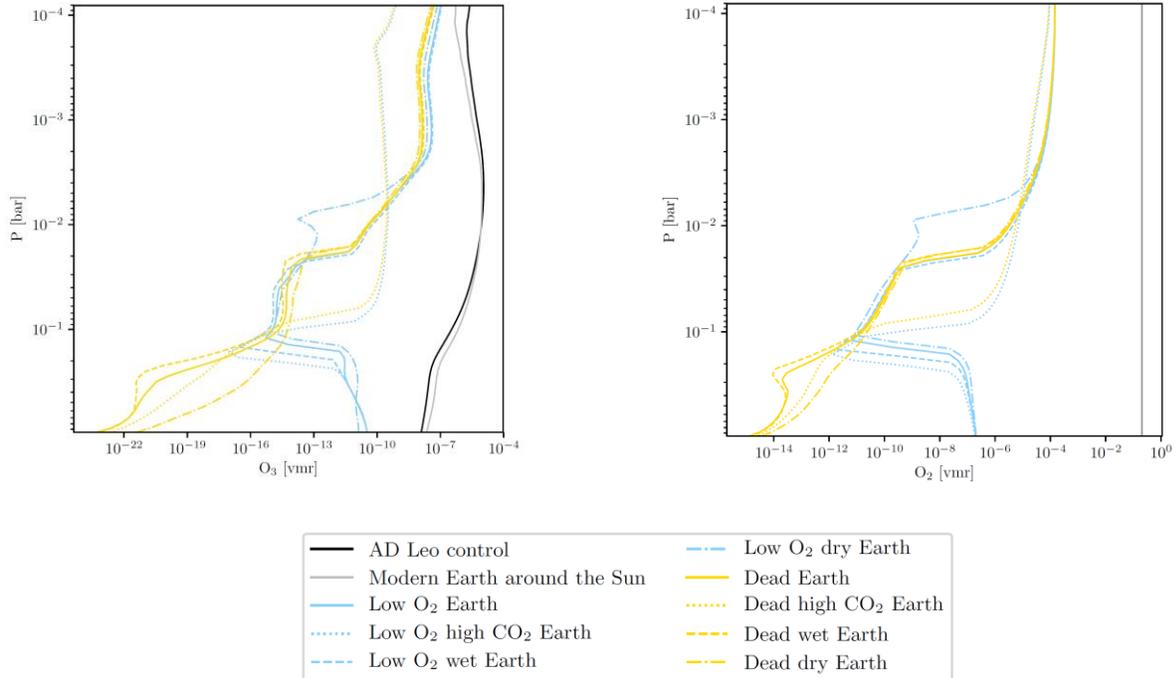

**Figure 1:** *$O_2$ (left) and $O_3$ (right) vmr profiles of various Earth-like scenarios orbiting in the HZ around the star AD Leo compared with the modern Earth around the Sun shown in grey. The other scenarios are grouped into 3 "families": "modern" Earth with Earth's biomass around AD Leo (solid black line, overlapped by grey line for $O_2$); an Earth with Earth's biomass and reduced $O_2$ concentrations (solid blue line) along with increased (100 PAL) $CO_2$ concentrations (short-dashed blue line), increased (100% sRH, 'wet') surface relative humidity (long-dashed blue line) and lowered (0.1% sRH, 'dry') surface relative humidity (dash-dotted blue line); an abiotic "dead" Earth with removed biomass (solid yellow line) along with increased $CO_2$ concentrations (short-dashed yellow line), increased surface relative humidity (long-dashed yellow line) and lowered surface relative humidity (dash-dotted yellow line). More details of the boundary conditions are shown in Tables 3 and 4.*

The "dead" Earth scenarios all cluster around $10^{-14}$ vmr $O_2$ (and $10^{-22}$ vmr $O_3$) in the lower atmosphere in Figure 1. Unlike in the low oxygen runs, $O_2$ in the "dead" Earth runs increases with increasing altitude in the lower atmospheric regions especially for the dry run (more so than the high $CO_2$ run). This was related to the diffusion of O produced by $CO_2$ photolysis from the upper layers down to the lower atmosphere, followed by the hydrogen oxides ($HO_x$) catalyzed production of $O_2$, which is consistent with the increase in $HO_x$ radicals towards the upper atmosphere (see Figure 6 (b)). For the high $CO_2$ cases (yellow dotted lines) the increased abundance of $CO_2$ molecules leads to stronger photolysis between 0.1 bar and 0.01 bar, hence increased $O_2$ vmr, whereas in the upper atmosphere smaller $O_2$ vmr are observed due to a higher vmr of the sink species H produced from a higher $H_2O$ abundance.

In summary, Figure 1 shows that interesting altitude-dependent chemical-climate couplings are appearing on separating the effects of lowering $O_2$ alone (blue curves) compared with lowering both $O_2$ as well as biomass emissions (yellow curves). Furthermore, large difference between the blue and yellow curves are apparent in the lower layers, but not in the upper layers which is for $O_3$ the region sampled by spectroscopic observations. This effect is due to the in-situ abiotic production in the upper layers e.g. due to $CO_2$ photolysis indicating that if life were not present (yellow lines), it would be hard to distinguish from planets with biospheres and low oxygen contents (blue lines) posing a possible "false-positive" detection when searching for life. In Figure 1 the responses of $O_3$ mostly follow those of $O_2$. Our results therefore suggest that $O_3$ is a good proxy for $O_2$ under the conditions investigated. However, the similar $O_3$ vmr for the yellow and blue curves in the upper layers also imply that it would be difficult to distinguish between abiotic or biological origins of $O_2$ (which forms $O_3$) via remote sensing measurements of $O_3$. Additional information such as the $O_2$ concentration would be informative in this respect although this is challenging as we show (see section 4.2 and 4.3).

### 4.1.2. Methane – $CH_4$

Figure 2 shows the $CH_4$ vmr profiles for the various scenarios. Firstly, the grey curve (representing the modern Earth around the Sun) suggests lower $CH_4$ vmr compared to the solid black curve which shows the AD Leo control case with modern Earth surface $CH_4$ emissions. This is consistent with increased output in the UVB region for the solar case compared with AD Leo which results in enhanced photolysis of $O_3$ for the modern Earth around the Sun scenario. This increases the reaction rates of the following chemical reactions:

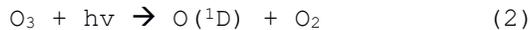

$$O_3 + h\nu \rightarrow O(^1D) + O_2 \qquad (2)$$

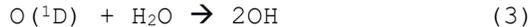

$$O(^1D) + H_2O \rightarrow 2OH \qquad (3)$$

which stimulates the production of the hydroxyl (OH) radical, the main sink for $CH_4$. This effect was also suggested by earlier model studies (see e.g. Segura et al., 2005). The black curve (AD Leo control run) exhibits an isoprofile of $10^{-3}$ vmr throughout the atmosphere, which indicates that $CH_4$ has a long dynamical lifetime and that its concentration is dominantly influenced by transport processes rather than chemical production and loss. Additionally, the incoming UVC radiation (not shown) which is an important $CH_4$ sink on the uppermost layers is more efficiently shielded in the high $O_2$ control run (black curve) compared to the lowered $O_2$ runs (blue and yellow curves), causing less photolytic destruction of $CH_4$ in the upper layers for the control case.

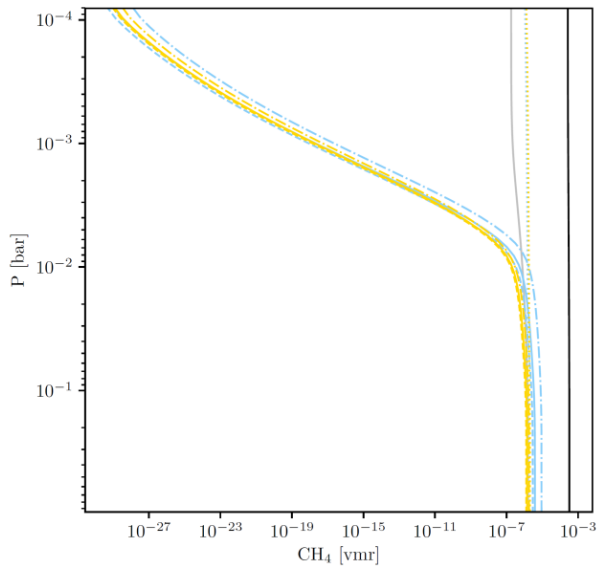

**Figure 2:** *As for Figure 1 but for $CH_4$.*

Secondly, the blue "family" of curves (lowered $O_2$ but other biomass fluxes as for modern Earth) clusters around $10^{-6}$ $CH_4$ vmr at the surface. Comparing chemical reaction rates for $CH_4$ in-situ sinks suggested that OH is one of the main sink species for $CH_4$ in the low to mid atmosphere via the reaction $CH_4 + OH \rightarrow CH_3 + H_2O$. Enhanced OH in the lower atmosphere was calculated for all blue curves compared to the black curve, i.e. AD Leo control run (see Figure 6 (d)). This was related to lowered $O_2$ leading to less UV shielding which generally favors OH (see chemical reactions (2) and (3)). The enhanced OH led to stronger $CH_4$ sinks for the blue line cases. There is also a sharp decrease in vmr around 0.01 bar for the blue curves which is related to $CH_4$ photolysis (with $\lambda < 120$ nm) as a leading sink reaction for all runs (except high $CO_2$) in the upper layers. The short-dashed blue (and red) curves (high $CO_2$ runs) in Figure 2 however indicate enhanced $CH_4$ in the upper layers. Further analysis suggested that this is due to the incoming stellar far-UV (FUV) radiation (at which $CH_4$ photolyzes) being shielded by the high $CO_2$ amount. Therefore, the leading $CH_4$ sink reaction for the high $CO_2$ runs is the reaction with OH, rather than photolysis which is the main sink for the other scenarios in the upper layers.

Thirdly, the yellow "family" of curves in Figure 2 (removed $O_2$ as well as biomass) also group in the lower atmosphere around $10^{-6}$ $CH_4$ vmr. This is rather unexpected, since the $CH_4$ biological surface flux has been removed for these runs, which would imply a lowering in vmr for the yellow curves compared to the blue curves which do have a biosphere producing $CH_4$. However, the geological sources from volcanic and metamorphic outgassing (see Table 2) are still present which help to maintain the atmospheric $CH_4$ value even without biomass emissions. Although the $CH_4$ biological surface flux in the "dead Earth" run is reduced to 1 molec./(cm$^2$ s) compared to the blue curves which have a $CH_4$ bio*logical surface flux* of $8.6 \cdot 10^{10}$ molec./(cm$^2$ s), the atmospheric $CH_4$ is only slightly reduced in the "dead" Earth run. This suggests a stabilizing effect (negative feedback) in which the initial change (a reduced surface flux of $CH_4$) is opposed by $CH_4$ photochemical effects in the atmosphere. Further analysis suggested that even though the flux for the "dead" Earth curves is reduced, $CH_4$ is nevertheless favored by reduced OH for the yellow curve scenarios compared to the blue curve scenarios (see Figure 6 (d)). This effect was related to $O(^1D)$ which was reduced, associated with a reduction in its photolytic precursors $O_2$ and $O_3$. Furthermore, $O(^1D)$ is an important species for the build-up of OH (via the reaction $H_2O$ + $O(^1D) \rightarrow 2OH$), which is a sink for $CH_4$. An important source of $O(^1D)$ in the lower atmosphere is $O_3$ photolysis ($O_3$ + hv $\rightarrow$ $O(^1D)$ + $O_2$. However, $O_3$ is reduced for the yellow curve cases, which suggests less $O(^1D)$), and therefore less OH. In summary, even though the $CH_4$ biological surface flux is reduced for the "dead Earth" scenarios, a strong reduction in $CH_4$ is nevertheless opposed by less OH. Overall, however, the blue line scenarios still have slightly higher $CH_4$ vmr due to biological sources on the surface. Note that in modern Earth's atmosphere, a positive (destabilizing) 'classical' feedback has been discussed (Prather, 1996) for $CH_4$ and OH. Here, increased $CH_4$ surface fluxes are an important sink for OH, which is reduced, enabling more $CH_4$ to build up, hence inducing a positive feedback. In our results, the 'lowered oxygen' (negative) feedback was stronger than the 'classical' (positive) feedback. In the upper layers a sharp decrease in $CH_4$ vmr can also be seen for the yellow curves. This is again related to $CH_4$ photolysis. The short-dashed yellow line (high $CO_2$ run) does not have a sharp decrease in vmr, consistent with the blocking of the incoming stellar radiation by the high $CO_2$ amount, preventing it from photolyzing $CH_4$.

In summary, Figure 2 shows that important feedbacks involving $CH_4$ can occur even when separating the effects of lowering $O_2$ and lowering both $O_2$ and biomass contents, especially near the surface where the atmospheric $CH_4$ content is favored by smaller amounts of the sink species OH. Note that more recent estimates of volcanic $CH_4$ fluxes suggested (Wogan et al., 2020) are substantially lower than those applied in this work. However, the maximum metamorphic $CH_4$ flux of $6.8 \cdot 10^8$ molec./($cm^2$ s) calculated by Guzmán-Marmolejo et al. (2013) could still lead to similar results as shown in Fig. 2.

### 4.1.3. Water — $H_2O$

The modeled $H_2O$ vmr profiles are shown in Figure 3. In the lower atmosphere (below the tropopause), the $H_2O$ concentrations are calculated in the convective-climate module since convection and the hydrological cycle are the relevant processes in this region. The short-dashed lines (high $CO_2$ content with 78% sRH) exhibit the highest $H_2O$ vmr due to greenhouse heating by $CO_2$ in the lower atmosphere and subsequent $H_2O$ evaporation. Contrastingly, the dashed dotted lines (0.1% sRH) exhibit the lowest surface vmr, grouping around $10^{-3}$ vmr. The long-dashed lines (100% sRH) have almost identical $H_2O$ vmr as the AD

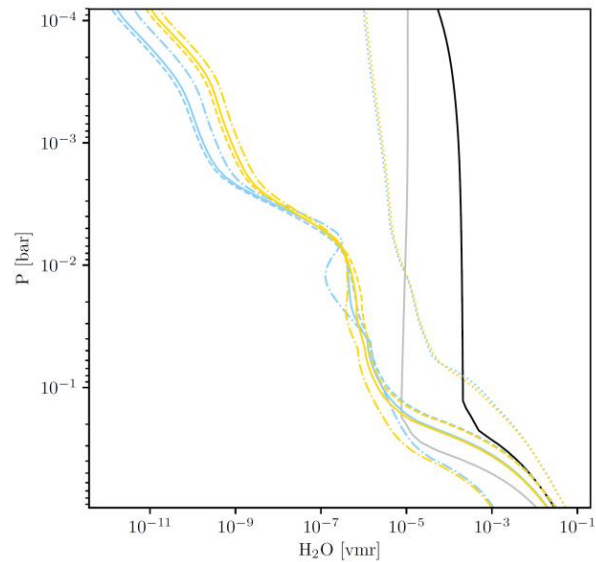

**Figure 3:** *As for Figure 1 but for $H_2O$.*

Leo control (78% sRH) black line, whereas the solid colored (yellow, blue and grey) curves have slightly lower vmr. This arises since the colored curves have reduced $CH_4$ (a greenhouse gas) contents in comparison to the black curve, as shown in Figure 2. Temperature and water vapor content decrease rapidly with altitude in the troposphere. Above this region, on the other hand, the chemistry module calculates the $H_2O$ vmr. Firstly, the reduced $CH_4$ abundances for the modern Earth around the Sun (see Figure 2) lead to less $H_2O$ production via $CH_4$ oxidation compared to the AD Leo control run, which exhibits approximately an isoprofile in the upper regions. This indicates that the $H_2O$ lifetime is high enough that it is not strongly influenced by chemical production and loss. The short-dashed yellow and blue curves (high $CO_2$ content) have enhanced $H_2O$ vmr. This is consistent with their large $CO_2$ content which shields the $H_2O$ molecules from being photolyzed. Thirdly, the yellow and blue "families" of curves also suggest a decreasing $H_2O$ vmr with height. Further analysis suggests that photolytic destruction by Lyman-alpha radiation is the dominating loss process together with a smaller contribution from the reaction $H_2O + O(^1D) \rightarrow 2OH$. The yellow curves have slightly higher $H_2O$ vmr compared to the blue curves in the upper regions which is mainly due to the enhanced production of $H_2O$ due to the reaction $OH + HO_2 \rightarrow H_2O + O_2$ (see Figure 6 (d)).

In summary, changing the relative humidity at the surface of the hypothetical Earth-like planets studied leads to clear compositional changes in the upper layers of the atmosphere compared to the black curve (AD Leo control run), even though the differences within the blue and yellow "families" of curves are mostly not strong. An exception is the short-dashed (high $CO_2$) curves where the high $CO_2$ content shields $H_2O$ from photolyzing in the upper atmosphere.

### 4.1.4. Nitrous Oxide — $N_2O$

Figure 4 shows the modeled $N_2O$ vmr profiles. Atmospheric sinks of $N_2O$ are mostly dominated by photolysis in the UV below 240 nm. The stronger UV environment for the modern Earth around the Sun (grey line) in the middle atmosphere is consistent with the reduced $N_2O$ vmr for this scenario compared to the black curve (AD Leo control run) which exhibits approximately an isoprofile around $10^{-6}$ vmr. This behavior indicates that atmospheric $N_2O$ abundance is dominated in the lower layers by the biological surface sources (mainly (de)nitrifying bacteria), has a long lifetime and is hence affected strongly by transport. Note that we do not consider chemodenitrification as a geological source of $N_2O$ (see e.g., Samarkin et al., 2010). Further analysis indicates that the main atmospheric sink for the AD Leo control scenario (black curve) is the reaction with $O(^1D)$, since the atmospheric profile of UVC (which photolyzes $N_2O$) is much weaker for the AD Leo control run due to its stronger $O_2$ shielding compared with other low oxygen runs.

Secondly, the "family" of blue curves (lowered $O_2$ contents) groups around $10^{-8}$ vmr at the surface, which is 100 times smaller compared to the AD Leo control black curve. This difference is mostly of photolytic nature. The flux of the UV radiation increases with height (not shown) and leads to stronger photolytic destruction of $N_2O$ especially in the upper layers and therefore a decrease in $N_2O$ vmr for pressures below ~0.1 bar.

Thirdly, the "family" of yellow curves (without surface biomass fluxes) in Figure 4 (with lowered $O_2$ and removed biomass) all group around $10^{-17}$ $N_2O$ vmr at the surface (with the exception of the short-dashed yellow line, which depicts high $CO_2$ contents). These very low vmr values arise due to the absence of a biomass (hence virtually no $N_2O$ surface emissions) for the yellow curves. These curves exhibit rather surprisingly an increase in vmr towards the upper atmosphere. Further analysis suggested this was due to enhanced abiotic production via the reaction $N_2 + O(^1D) \rightarrow N_2O$ on the upper layers. An increase of $O(^1D)$ with height is shown in Figure 6 (c). To be noted is that the short-dashed yellow curve (high $CO_2$) in Figure 4 stands out compared to other yellow curves. Further analysis suggested that the hypothetical

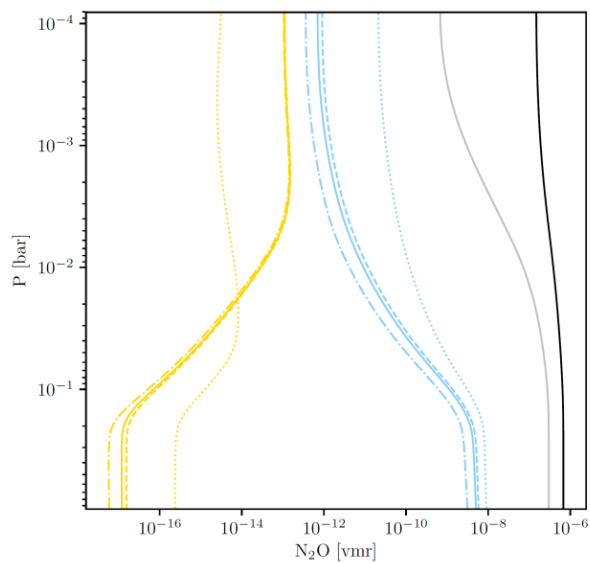

**Figure 4**: *As for Figure 1 but for $N_2O$.*

Earth in this scenario experiences photolytic shielding due to high $CO_2$ contents, which hinders UV radiation from producing $O(^1D)$. Since the $O(^1D)$ atoms can act as a sink for $N_2O$ in the lower layers, but as an abiotic source for $N_2O$ in the upper layers of the atmosphere, the reduced abundance of $O(^1D)$ atoms due to $CO_2$ shielding is consistent with the reduced (increased) $N_2O$ vmr of the high-$CO_2$ yellow curve in the upper (lower) layers compared to the other yellow curves.

In summary, despite removing the biomass at the surface for the yellow curve scenarios, the atmosphere still produces $N_2O$ abiotically in the upper layers of the atmosphere so that its vmr here is similar to that of the blue curve scenarios in which biomass is present. This process could create a "false-positive" detection by remote observations when searching for life. The high-$CO_2$ scenarios, however, exhibit this behavior to a lesser degree in the upper regions, and in this sense $N_2O$ for these cases could be considered a better biosignature.

### 4.1.5. Chloromethane – $CH_3Cl$

The modeled $CH_3Cl$ vmr profiles are shown in Figure 5. Firstly, the loss processes of $CH_3Cl$ in the middle atmosphere are dominated by photolysis in the UVC. The stronger UVC environment for the modern Earth around the Sun (grey line) as already discussed is consistent with the reduced $CH_3Cl$ vmr for this scenario compared to the black curve (AD Leo control run) which features $CH_3Cl$ concentrations close to an isoprofile of around $10^{-7}$ vmr with some reduction in the upper layers. This indicates that for the AD Leo control run $CH_3Cl$ has a relatively long lifetime and that its concentration profile is mainly influenced by transport processes rather than chemistry.

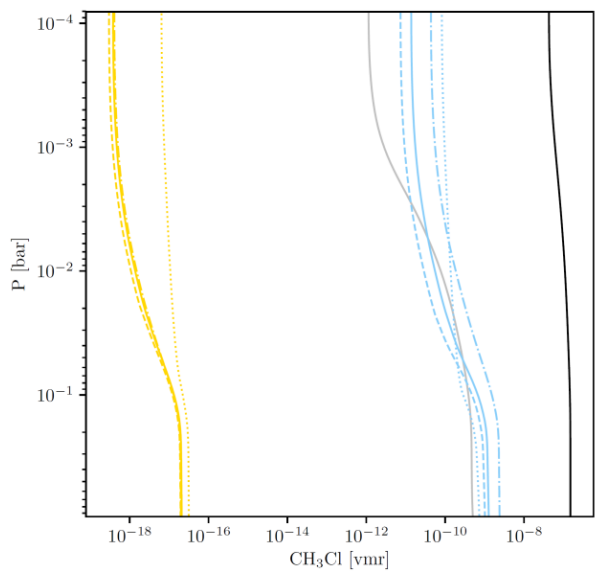

**Figure 5**: *As for Figure 1 but for CH₃Cl.*

Secondly, the blue "family" of curves (lowered $O_2$ content but with biomass) groups around $10^{-9}$ vmr at the surface. The difference between the blue curves and the black curve in the lower layers is consistent with a much weaker ozone layer for the low oxygen blue curves, which enables incoming stellar UV radiation to penetrate deeper into the atmosphere and photolyze more $CH_3Cl$. Moreover, the blue lines have a higher concentration of the sink species OH compared to the black line (as shown in Figure 6 (d)), which is consistent with the lower $CH_3Cl$ vmr of the blue curves. Furthermore, a decrease in vmr with height can be seen for the blue lines, associated with increasing photolytic destruction of $CH_3Cl$ on the upper levels.

Thirdly, the yellow "family" of curves in Figure 5 (removed $O_2$ and biomass) all group around $10^{-17}$ vmr at the surface. Furthermore, the vmr of $CH_3Cl$ decrease with height for the yellow lines, consistent with the increase in the photolysis rates due to stronger UV. The dotted yellow line (high $CO_2$ content) in Figure 5 stands out compared to the other yellow lines, as it exhibits an (almost) isoprofile throughout the atmosphere. This is due to the high $CO_2$ content shielding $CH_3Cl$ from photolytic destruction in the upper layers of the atmosphere.

In summary, atmospheric $CH_3Cl$ displays a decrease with decreasing surface emissions. Furthermore, the results suggest that changing the $CO_2$ and sRH affect only modestly the $CH_3Cl$ vmr. Moreover, $CH_3Cl$ can be considered a good biosignature in the sense that there is a clear separation between the yellow and blue curves in the upper layers of the atmosphere, where spectroscopic measurements will likely be taken.

### 4.1.6. Minor species

**(a) Nitrogen Oxides – $NO_x$**

Figure 6 (a) is as Figure 5 but for $NO_x$. The black curve (AD Leo control run) shows a slight increase in the lower layers of the atmosphere, featuring a minimum between 0.1 and 1.0 bar. The modern Earth around the Sun (grey line) exhibits increased $NO_x$ vmr. The yellow and blue "families" of curves also exhibit a slight increase in $NO_x$ vmr with height in the lower layers of the atmosphere, followed by a decrease in the middle layers between ~0.3 bar and 0.01 bar. In general, interpreting such results depends on potentially complex interplays between the ability of reservoir molecules to release $NO_x$ at a given altitude, which depends on the properties of the particular reservoir molecule, the availability of UV and for some reservoirs (like $N_2O_5$) on the local temperature. On Earth, $NO_x$ peaks in the stratosphere due to its photolytic release from e.g. $N_2O$ and $HNO_3$, which is mostly reproduced in Figure 6(a) (grey line). In the upper atmosphere, $NO_x$ behaves mostly as an isoprofile in all scenarios. In the lower layers, $NO_x$ is related to $O_3$ formation via the smog mechanism, which requires $O_2$, volatile organic compounds, nitrogen oxides, and UV radiation. Increases in $NO_x$ vmr in the lower atmosphere correspond with $O_3$ production for the yellow lines. Furthermore, all scenarios show considerable local structure in the vertical, related to strong coupling between $NO_x$ and $HO_x$ families (see Figure 6 (b)). In general, $NO_x$ is produced in the troposphere through lightning and surface emissions, and is removed by washout.

**(b) Hydrogen Oxides – $HO_x$**

Figure 6 (b) is as Figure 5 but for $HO_x$. The results suggest that $HO_x = (OH + HO_2 + H)$ generally increases towards lower pressures. The photolytic destruction of H$_2$O by Lyman-alpha radiation plays a dominant role in the production of HO$_x$ in the upper layers. To be noted is that the blue and yellow "families" of curves almost always exhibit larger HO$_x$ vmr compared to the black and grey curves (AD Leo control run and modern Earth around the Sun). Further analysis suggests that this is related to the UV photons which penetrate deeper into the atmosphere for the yellow and blue curves which have lower O$_2$ and hence less UV shielding compared to the black curve, thus favoring $H_2O$ photolysis and consequently producing larger amounts of $HO_x$.

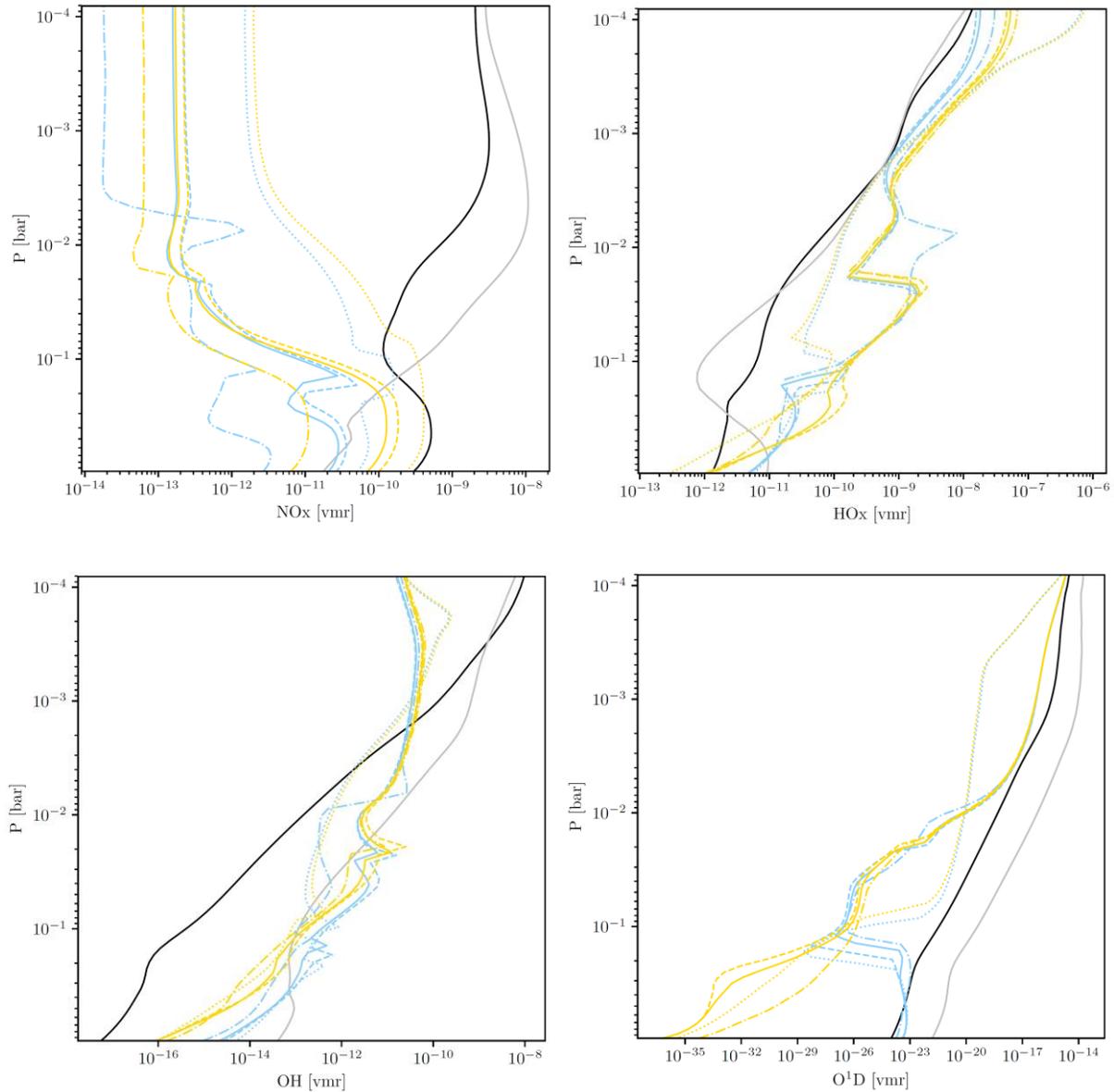

**Figure 6:** *As for Figure 1 but for NO$_x$ (upper left), HO$_x$ (upper right, O($^1$D) (lower left) and OH (lower right).*

**(c) Vibrationally-excited atomic oxygen – O($^1$D)**

Figure 6 (c) is as Figure 5 but for O($^1$D). This species shows a similar vertical structure as $O_2$ and $O_3$ for the yellow and blue "families" of curves, as it is directly produced from those species via photolysis (see Figure 1). The grey line (modern Earth around the Sun) exhibits the highest O($^1$D) vmr, consistent with its increased $O_3$ photolysis due to its stronger UVC environment compared to the black line. Furthermore, O($^1$D) is important for the production of OH via the reaction $H_2O$ + O($^1$D) → 2OH, which is a sink for $CH_4$.

**(d) Hydroxyl radical – OH**

Figure 6 (d) is as for Figure 5 but for OH. The black curve (AD Leo control run) exhibits a steady increase in vmr towards higher altitudes. Further analysis suggests that this is related to the increased vmr of O($^1$D) due to enhanced $O_3$ photolysis (see Figure 6 (c)), which partakes in the production reaction $H_2O$ + O($^1$D) → 2OH. Increased OH vmr for the modern Earth around the Sun (grey line) compared to the black line is consistent with its stronger UV environment responsible for $O_3$ photolysis. Moreover, another source of OH is via $H_2O$ photolysis ($H_2O$ + hv → OH + H). Since the FUV radiation of the black curve scenario does not penetrate as deep into the atmosphere compared to the blue and yellow curves, less OH is produced via photolytic destruction of $H_2O$ compared to the colored lines.

**4.1.7. Temperature Profiles**

Figure 7 summarizes all temperature-pressure profiles calculated in our model study. The AD Leo control scenario (solid black curve) in Figure 7 features ~15K surface warming compared to the Earth around the Sun control (grey line) due to $CH_4$ greenhouse heating (see e.g. Segura et al., 2005). The AD Leo control scenario (black curve) displays the familiar temperature decrease from the surface up to around 0.1 bar due to the imposed adiabatic lapse rate. The rate of cooling weakens with decreasing pressure from 0.1 to $10^{-3}$ bar due to the modest $O_3$ layer (see Figure 1), which leads to some $O_3$ heating. However, the $O_3$ layer of the black curve in Figure 1 is smaller than that of the modern Earth around the Sun (grey line) which exhibits a noticeable temperature inversion. This is mainly due to the weaker UV flux of the M-star AD Leo compared to the Sun for wavelengths longer than 200 nm, which is the wavelength range responsible for radiative heating due to the absorption of UV radiation by $O_3$ molecules. For altitudes above the aforementioned pressure region, $O_3$ is scarce and the temperatures decrease more strongly with height. The blue and yellow "families" of curves display broadly similar temperature behavior. The largest differences arise on changing the $CO_2$ and sRH. Compared to the AD Leo control run (black solid line), all other runs (except for the long-dashed enhanced $CO_2$ curve) have lower surface temperatures. Further analysis suggested that this is due to a decrease in the greenhouse gases $H_2O$ and $CH_4$ as previously discussed above. The yellow and blue dotted lines, however, have enhanced $CO_2$ and hence increased greenhouse warming in the lower regions. Additionally, they exhibit a local minimum in temperature between $10^{-2}$ bar and $10^{-3}$ bar mainly due to $CO_2$ cooling. The AD Leo control scenario is approximately 40 K warmer than the other runs (except the high $CO_2$ runs) in the middle atmosphere. Further analysis suggests that this is mainly due to a local $O_3$ minimum (see Figure 1), and thus less radiative heating for the blue and yellow scenarios. This minimum occurs in the region with the corresponding minimum in $O_2$ (see Figure 1), caused by a maximum in $HO_x$ as previously discussed (see Figure 6 (b)). Furthermore, there is a weak temperature inversion for the blue and yellow curves in the middle atmosphere region at around $10^{-2}$ bar, respectively. Further analysis suggested that this weak inversion is mainly due to two processes: firstly due to enhanced photolytic destruction of $CO_2$ and thus less $CO_2$ cooling compared with the AD Leo control (black curve), and secondly due to more $O_3$ heating in this region due to a higher vmr of $O_2$ (hence $O_3$) (see Figure 1) produced by $CO_2$ photolysis followed by $HO_x$ catalyzed recombination of O to form $O_2$, and thus $O_3$. In summary, changing the sRH and $CO_2$ contents of a hypothetical planet around the M-dwarf star AD Leo influences the temperature-pressure profile more strongly than changing the $O_2$ and biomass contents (blue and yellow curves).

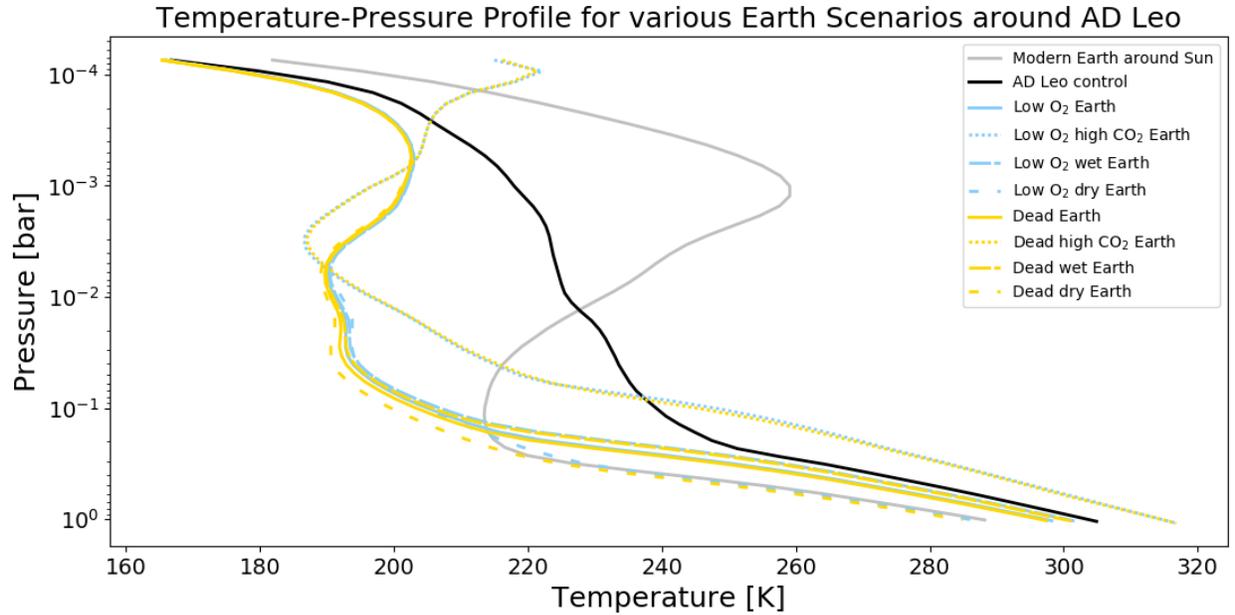

**Figure 7**: *Temperature vs. pressure profiles of various Earth-like scenarios orbiting in the HZ around the star AD Leo. The modern Earth around the Sun temperature profile (grey line) is also plotted for comparison. The scenarios (Tables 1,3 and 4) are grouped into 3 "families": (1) The AD Leo control with Earth's biomass (solid black line); (2) Earth's biomass and reduced $O_2$ concentrations (solid blue line) along with increased $CO_2$ concentrations (short-dashed blue line); increased surface relative humidity (long-dashed blue line) and lowered surface relative humidity (dash-dotted blue line); (3) an abiotic "dead" Earth with removed biomass (solid yellow line) along with increased $CO_2$ concentrations (short-dashed yellow line), increased surface relative humidity (long-dashed yellow line) and lowered surface relative humidity (dash-dotted yellow line). The boundary conditions are shown in Tables 1-4.*

## 4.2. Transmission (Effective Height) Spectra

Table 5 summarizes relevant absorption wavelengths of important atmospheric constituents. The transmittance spectra for the various scenarios of hypothetical Earth-like planets around AD Leo are shown in Figure 8. The black solid line represents the AD Leo control run, whereas the yellow and blue lines denote the "dead" and low oxygen Earth runs around AD Leo, respectively. The main features of Figure 8 include a Rayleigh scattering feature in the UV and visible region below 1 μm, a large number of $CH_4$ and $H_2O$ rotational-vibrational bands in the (near) IR, the strong $CO_2$ fundamental bands around 4.3 and 15 μm and various biosignature bands e.g. for $O_2$ (~0.76 μm), $N_2O$ (near IR), the $O_3$ Hartley-Huggins band (0.31-0.34 μm), $O_3$ Chappius band (between 0.4 and 0.85 μm) and the $O_3$ fundamental band at ~9.6 μm. The largest effective heights range from several tens of km for the AD Leo control run down to a few km for the "dead Earth" runs. Furthermore, the AD Leo control run has higher effective heights for almost all atmospheric constituents, apart from the two high $CO_2$ runs, whose high concentrations of

| Species | Wavelength [μm] | Reference |
|---------|-----------------|-----------|
|         |                 |           |

| O₂ | 0.63, 0.69, 0.76, 0.77, 0.86, 1.06, 1.27, 1.58 | Catling et al. (2017) |
|---|---|---|
| O₃ | 0.31-0.34,0.4-0.85, 4.8, 9.6 | Serdyuchenko et al. (2014) Brunetti & Prodi (2015) |
| CH₄ | 0.9, 1.1, 1.4, 1.7, 2.2, 2.3, 3.3, 6.5, 7.1, 7.7 | Brunetti & Prodi (2015) |
| H₂O | 0.82, 0.94, 1.1, 1.4, 1.9, 2.7, 5.5-7.0, >27 | Solomon et al. (1998) |
| N₂O | 2.8, 3.5, 3.9, 4.4, 7.5 | Marshall & Plumb (2007) |
| CO₂ | 1.4, 1.6, 2.0, 2.7, 4.3, 4.8, 9.5, 10.5, 15.0 | Catling et al. (2017) |
| CH₃Cl | 3.3, 7.0, 9.7, 13.7 | Schwieterman et al. (2018) |
| CO | 1.6, 2.4, 4.7 | Brunetti & Prodi (2015) |

**Table 5:** *Relevant absorption wavelengths of important atmospheric constituents in Earth-like atmospheres.*

$CO_2$ block the stellar shortwave (UV) radiation and therefore increase the effective heights at most wavelengths compared to the AD Leo control run. In the middle atmosphere, $H_2O$ is more abundant than $CH_4$ (see Figures 3 and Figure 2) for all runs, which leads to the absorption features at 1.1 µm and 1.4 µm being dominated by $H_2O$. Furthermore, a large spectral features of $O_3$ can be seen below 1 µm and at 9.6 µm (black line) since AD Leo is an active M-dwarf star whose strong FUV radiation photolyzes $CO_2$ and $O_2$ efficiently, resulting in an increased $O_3$ abundance compared with less active M-dwarf stars. However, the 9.6 µm $O_3$ feature is masked by $CO_2$ in all other scenarios. Only the AD Leo control run indicates $O_2$ features in the visible region at 0.68 µm and 0.76 µm (see Table 5 for reference). The so-called "Rayleigh-slope" due to atmospheric scattering is also evident below 1 µm in all runs, which could in principle enable the determination of the planet's atmospheric scale height and hence (given temperature) the bulk atmospheric constituents in the absence of hazes. In order to derive the impact of the $O_2$ collision induced absorption (CIA) at 6.4 microns as applied by Fauchez et al. (2020) upon our calculated spectra, we performed a sensitivity study where we considered the $O_2$-$O_2$ and $O_2$-$N_2$ CIAs in GARLIC. The $O_2$-$O_2$ CIA is taken from HITRAN 2016 (Karman et al., 2019) and for the $O_2$-$N_2$ CIA we assume the same absorption efficiency as for the $O_2$-$O_2$ CIA (following Fauchez et al., 2020, and Rinsland et al., 1989). In our simulated scenarios the impact upon the effective height spectra is negligible due to stronger absorption by $H_2O$ and $CH_4$ around 6.3 microns compared to Fauchez et al. (2020).

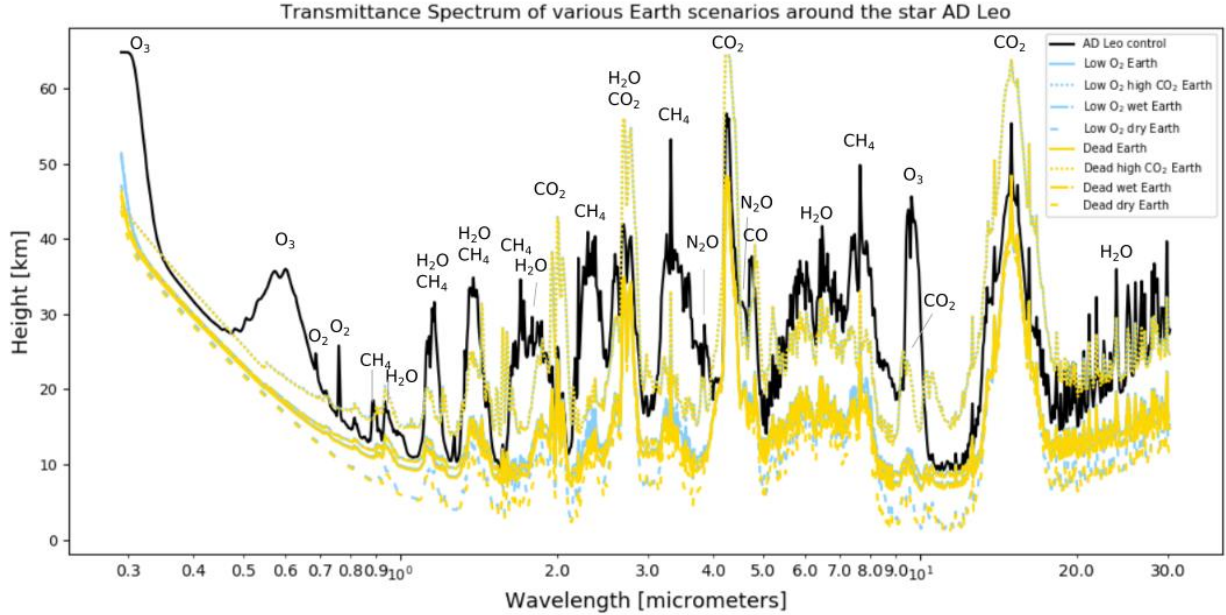

**Figure 8:** *Transmission (effective height) spectra of various Earth-like scenarios around the M-dwarf star AD Leo in effective height. The scenarios are grouped into 3 "families": AD Leo control run with Earth's biomass (solid black line); an Earth with Earth's biomass and reduced $O_2$ concentrations (solid blue line) along with increased $CO_2$ concentrations (short-dashed blue line), increased surface relative humidity (long-dashed blue line) and lowered surface relative humidity (dash-dotted blue line); an abiotic "dead" Earth with removed biomass (solid yellow line) along with increased $CO_2$ concentrations (short-dashed yellow line), increased surface relative humidity (long-dashed yellow line) and lowered surface relative humidity (dash-dotted yellow line). Further information on the boundary conditions is provided in Tables 3 and 4.*

### 4.2.1. Transmission (Effective Height) Difference Spectra

Figure 9 shows effective height difference spectra[6] for (1) the low oxygen minus the "dead Earth" scenarios (top panel), (2) "dead Earth" scenarios (with changed $CO_2$ and sRH) minus "dead Earth" (middle panel) and (3) low oxygen (with changed $CO_2$ and sRH) minus low oxygen scenarios (bottom panel) scenarios. The upper panel shows the largest spectral differences in $CH_4$ spectral features at 1.7 µm, 2.3 µm, 3.3 µm and 7.7 µm (see Table 5 for reference) (orange line) of 5-6 km for low oxygen dry minus dead Earth dry due to the presence of a methanogenic biosphere, leading to strong, positive values for the difference in effective heights. For more moist scenarios (magenta and blue lines, upper panel of Figure 9) $CH_4$ is reduced e.g. by OH and the spectral differences are reduced by a factor of 2 compared with the orange line. The high $CO_2$ scenarios (green line, upper panel of Figure 9) (with even more moist tropospheres) show the smallest differences in effective heights, with a few spectral differences evident in the $CH_4$ bands at 2.3 µm and 3.3 µm. A pronounced difference in effective height of 3-5 km between low $O_2$ and "dead" Earth is observed for the $O_3$ Hartley-Huggins band at 0.3 µm except for the high $CO_2$ scenarios.

The middle panel of Figure 9 shows the largest differences in effective heights of more than 20 km for the "dead" Earth with high $CO_2$ contents minus the "dead" Earth run (magenta line), at the $CH_4$ (1.1 µm, 1.4 µm, 1.7 µm and 7.7 µm), $CO_2$ (2.0 µm, 2.8 µm, 4.3 µm and approx. 15 µm) and $H_2O$ (1.1 µm, 1.4 µm, and 2.8 µm) bands. The double $CO_2$ feature at 9.5 µm and 10.5 µm, which masks the $O_3$ band at 9.6 µm, can also be seen in the magenta difference spectrum. The "dead" high $CO_2$ run has a high $H_2O$ content compared to the "dead" Earth run (see Figure 3), which leads to the strong differences in $H_2O$

---

*The spectra in the top panel of Figure 9 are obtained by subtracting the "dead" Earth transmission spectrum from the reduced $O_2$ with biomass Earth transmission spectrum (magenta line); subtracting the "dead" enhanced $CO_2$ Earth transmission spectrum from the reduced $O_2$ enhanced $CO_2$ with biomass Earth transmission spectrum (green line); subtracting the "dead" enhanced sRH Earth transmission spectrum from the reduced $O_2$ enhanced sRH with biomass Earth transmission spectrum (blue line) and subtracting the "dead" Earth reduced sRH transmission spectrum from the reduced $O_2$ and sRH with biomass Earth transmission spectrum (orange line).*

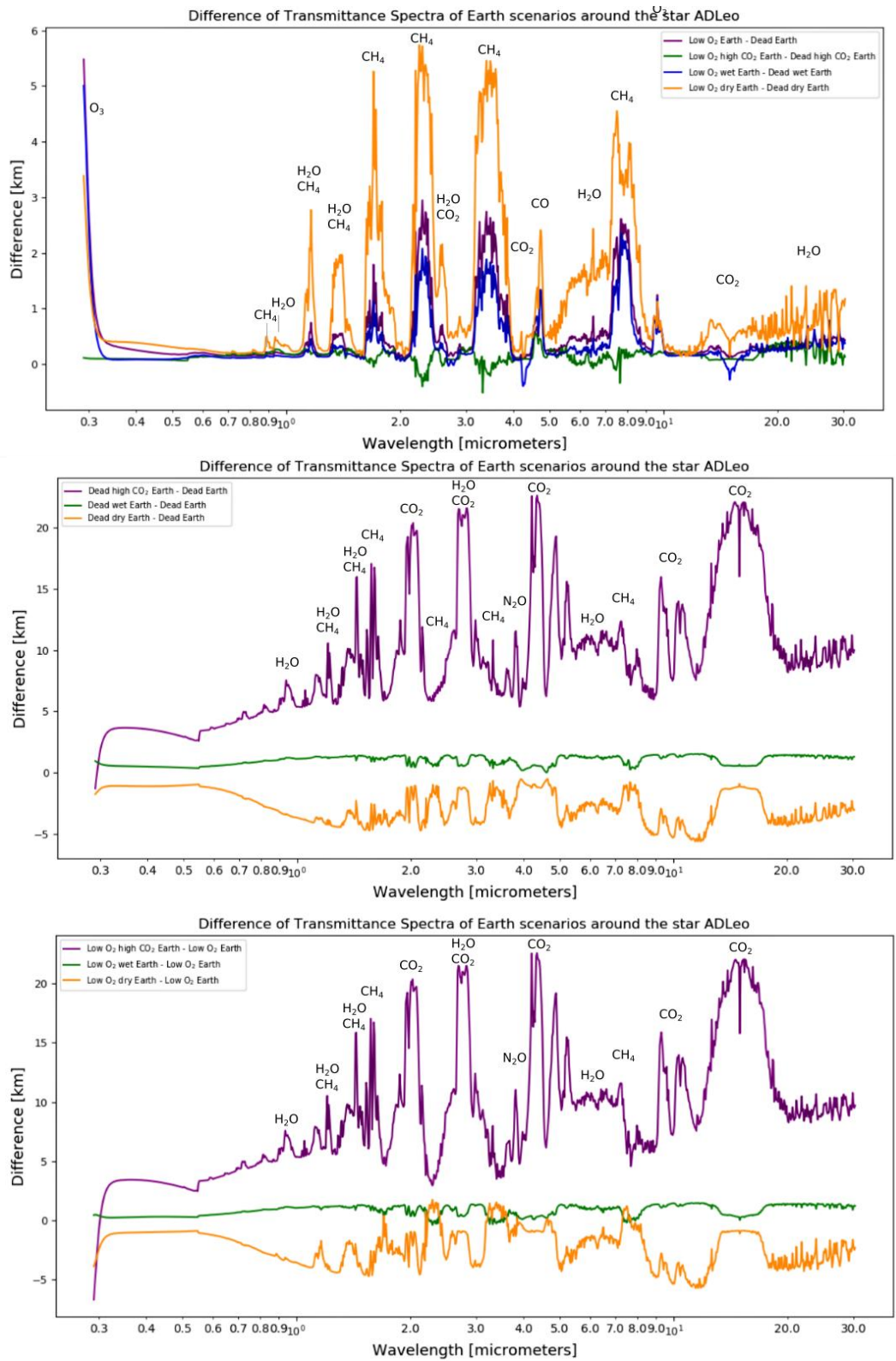

**Figure 9:** *Difference spectra of various Earth scenarios around the star AD Leo. Upper panel: the spectra are obtained by subtracting the "dead" Earth*

*transmission spectrum from the reduced $O_2$ Earth transmission spectrum. Middle panel: spectra obtained by subtracting the "dead" Earth transmission spectrum from the "dead" enhanced $CO_2$ (magenta line), "dead" enhanced sRH (green line) and "dead" reduced sRH (orange line) Earth transmission spectrum, respectively. Lower panel: spectra obtained by subtracting the reduced $O_2$ Earth transmission spectrum from the reduced $O_2$ enhanced $CO_2$ (magenta line), reduced $O_2$ enhanced sRH (green line) and reduced $O_2$ and sRH (orange line) Earth transmission spectrum, respectively.*

bands for the magenta line. The "dead" high $CO_2$ runs have higher $CH_4$ and $CO_2$ concentrations compared to the "dead" Earth run (see Figure 2), which is also evident in the magenta line. A small difference in effective heights can also be seen at the 3.9 μm $N_2O$ band for the magenta line. The green line (middle panel, Figure 9) shows the smallest differences in this panel. Furthermore, the orange line in the middle panel also shows rather weak negative differences, indicating that the continuum of the "dead" Earth scenario has slightly higher effective heights than that of the "dead" dry Earth scenario.

The bottom panel of Figure 9 shows the differences for the various low oxygen Earth runs. These differences are similar in magnitude to those between the "dead" Earth runs (middle panel), with the major difference of more than 20 km arising between the low oxygen high $CO_2$ Earth and the low oxygen Earth (magenta line).

In summary, changing the biomass produces only a modest difference in effective height of up to ≈ 6 km (see top panel of Figure 9), whereas changing the $CO_2$ content produces large differences of several tens of km in effective height (up to 25 km, as shown in the middle and bottom panels of Figure 9). Thus, the latter effects produce larger differences in the transmission spectrum of a hypothetical Earth-like planet in the HZ of the M-dwarf star AD Leo.

### 4.2.2. $O_2$ Parameter Study

In order to derive the minimum necessary atmospheric $O_2$ concentration for the $O_2$ features to be evident in a transmission spectrum of a hypothetical Earth-like planet orbiting in the HZ of the M-dwarf star AD Leo, an $O_2$ parameter study was conducted. For this, the present atmospheric level (PAL) of 21% $O_2$ (black line in Figure 10), as well as consecutively decreasing $O_2$

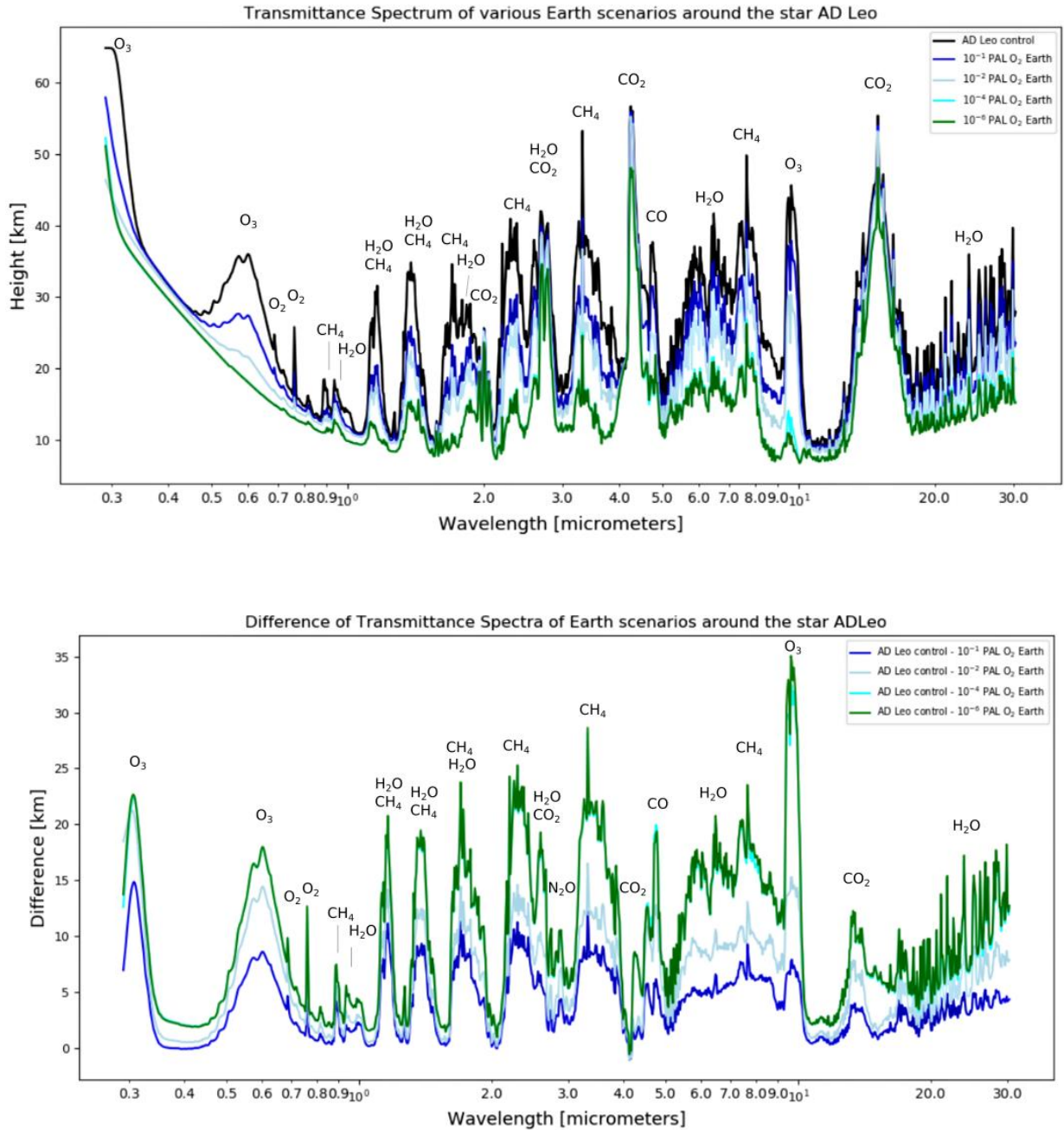

**Figure 10**: *Top panel: Transmission spectra of hypothetical Earth-like planets around the M-dwarf star AD Leo with Earth's biomass and varying O₂ contents – "modern" Earth (black line), 0.1 PAL O₂ (dark blue line), 10⁻² PAL O₂ (light blue/grey line), 10⁻⁴ PAL O₂ (cyan line) and 10⁻⁶ PAL O₂ (green line). Bottom panel: Difference transmission spectra obtained by subtracting the transmission spectrum of a hypothetical Earth with Earth's biomass and varying O₂ contents – 0.1 PAL O₂ (dark blue line), 10⁻² PAL O₂ (light blue/grey line), 10⁻⁴ PAL O₂ (cyan*

*line) and $10^{-6}$ PAL $O_2$ (green line) from the transmission spectrum of the AD Leo control run with 21% $O_2$ (1 PAL).*

concentrations of 0.1 PAL $O_2$ (dark blue line), $10^{-2}$ PAL $O_2$ (light blue/grey line), $10^{-4}$ PAL $O_2$ (cyan line) and $10^{-6}$ PAL $O_2$ (green line) were chosen. All scenarios contain Earth's biomass and only the $O_2$ contents have been changed. Figure 10 shows the transmission spectra[7] of the $O_2$ parameter study scenarios of hypothetical Earth-like planets orbiting in the HZ of the M-dwarf star AD Leo in the top panel, as well as the difference spectra of the AD Leo control run and the respective $O_2$ scenarios in the bottom panel. The results suggest that lowering the $O_2$ content decreases the effective heights of most of the prominent spectral features, like those of $CH_4$, $H_2O$, $O_3$ and $O_2$. The $CO_2$ bands at 2.0 μm, 2.7 μm, 4.3 μm and 15 μm do not seem to be greatly influenced by the varied $O_2$ content, as can also be seen in the difference spectrum in the bottom panel. For an $O_2$ concentration of 0.1 PAL (dark blue line) the effective height at 0.77 μm is only approximately half of that of the AD Leo control run (black line) (see Figure 10 top panel). The scenarios with even less $O_2$ are barely visible in the transmission spectrum. The $O_3$ features below 1 μm and at 9.6 μm is clearly evident for $O_2$ concentrations of $10^{-2}$ PAL and higher. The results also suggest that the largest difference in the 0.77 μm $O_2$ feature occurs between the 1 PAL (black line) and the $10^{-6}$ PAL $O_2$ scenario, with a difference of approximately 10 km. The strongest differences are apparent for both the $10^{-6}$ PAL $O_2$ (green line) and the $10^{-4}$ PAL $O_2$ (cyan line). The largest differences in the 9.6 μm $O_3$ band are also apparent between the $10^{-6}$ PAL $O_2$ (green line) or $10^{-4}$ PAL $O_2$ (cyan line) and the AD Leo control run of approximately 35 km. Interestingly, this difference decreases to 15 km for $O_2$ concentrations of $10^{-2}$ PAL (light blue/grey line) and diminishes to about 5 km for $O_2$ contents of 0.1 PAL (dark blue line). In summary, the $O_2$ feature is only visible at 0.77 μm in transmission spectroscopy for $O_2$ concentrations of 0.1 PAL $O_2$ (blue line) and 1

---

The difference spectra in the bottom panel of Figure 10 are obtained by subtracting the transmission spectrum of a hypothetical Earth-like planet with Earth's biomass and the respective $O_2$ concentration ($10^{-1}$ PAL, $10^{-2}$ PAL, $10^{-4}$ PAL or $10^{-6}$ PAL) from the transmission spectrum of "modern" Earth with 1 PAL $O_2$.

PAL (black line). The O₃ features in transmission spectra are clearly visible for O₂ concentrations of $10^{-2}$ PAL (light blue/grey line) and higher.

### 4.3. Detectability of CH₄ and O₂ spectral signals

Figure 11 shows the number of transits required to detect CH₄ and O₂ using ELT HIRES (high resolution spectroscopy) with $S/N$ = 5. The dashed line is the position of AD Leo. Table 6 shows the number of transits at 5 pc required to detect atmospheric spectral features of O₂ and CH₄ for the simulated planets with AD Leo-type instellation. For CH₄, Figure 11 and Table 6 suggest that for all runs CH₄ is detectable with less than 11 transits at the position of AD Leo (4.9 pc).

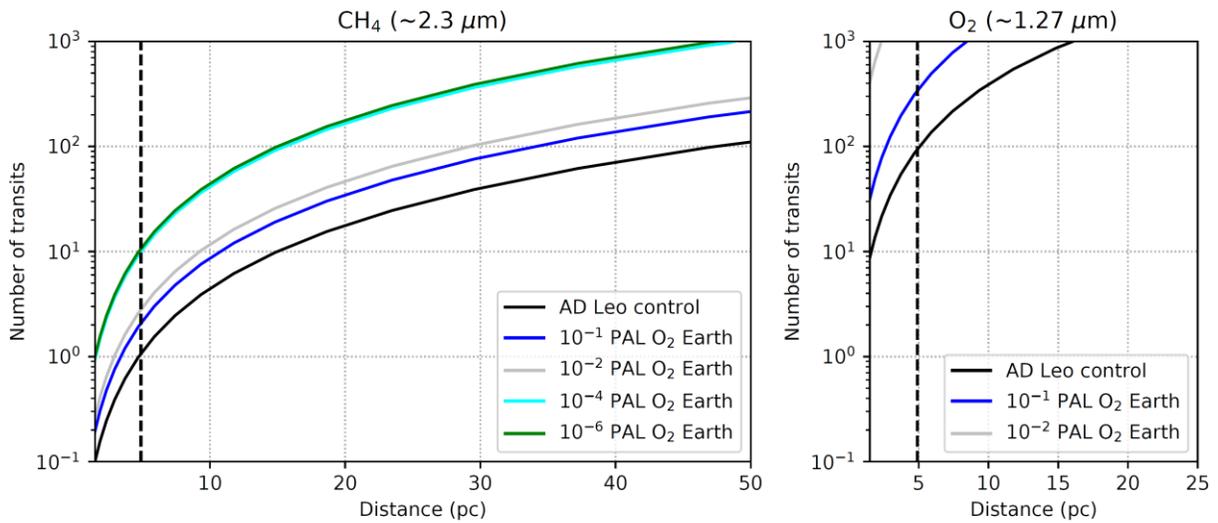

**Figure 11:** *Number of transits required to detect (a) CH₄ up to 50 pc (left panel) and (b) O₂ up to 25 pc (right panel). The vertical dashed line shows distance = 5 pc. Solid lines shown in the legend indicate results for the model scenarios with varying oxygen amounts which are described in section 3.3.*

The lowest number of transits is obtained for the AD Leo control run (1 PAL O₂). CH₄ is detectable with less than 10 transits up to 15 pc for the AD Leo control run, whereas it is difficult to detect for the $10^{-4}$ and $10^{-6}$ PAL O₂ scenarios because high UV leads to strong CH₄ removal in the upper atmosphere. By comparison, similar calculations, but for the James Webb Space Telescope (JWST) (Wunderlich et al., 2019), suggested that ~10 transits would be required for such a system at 10 pc. At 4.9 pc AD Leo is at the saturation limit for JWST.

| Oxygen level (PAL) | $N_{transits}$ for $O_2$ | $N_{transits}$ for $CH_4$ |
|---|---|---|
| 1 | 94 | 1.07 |
| $10^{-1}$ | 339 | 2.08 |
| $10^{-2}$ | 4454 | 2.81 |
| $10^{-4}$ | $8.96 \cdot 10^6$ | 10.6 |
| $10^{-6}$ | $5.30 \cdot 10^7$ | 10.7 |

**Table 6:** *Number of transits ($N_{transits}$) at 5 pc for [S/N=5.0] detection of (a) $CH_4$ (2.3μm) and (b) $O_2$ (1.27μm).*

For $O_2$, Figure 11 and Table 6 suggest that 94 transits are required to detect an Earth-like planet with a 1 PAL $O_2$ atmosphere orbiting in the HZ of AD Leo at 4.9 pc distance, hence a detection of even an Earth-like oxygen rich atmosphere is already challenging. For scenarios with an $O_2$ content of 0.1 PAL or lower, the detection of $O_2$ for the 1.27 μm band is favored compared to the 0.76 μm band (not shown), which is different to the results of Rodler & López-Morales (2014) who suggest that the observation of the $O_2$ at 0.76 μm band is more favored for a planet around an M3 star. Our results suggest that although the atmospheric signal is indeed improved at 0.76 μm, the photon flux is lower and the instrument noise is expected to be larger in the visible.

## 5. Summary and Conclusions

Our work has investigated the response and detectability of the atmospheres of hypothetical Earth-like planets (1 Earth mass and radius, placed at a distance of 0.153 AU from the host star AD Leo, in order to receive a total top-of-atmosphere energy of 1367 W/m$^2$ corresponding to the mean instellation which the modern Earth receives from the Sun at 1 AU) upon reducing $O_2$ concentrations, removing biomass emissions, and varying $CO_2$ content and surface relative humidity (sRH). The scenarios were performed using the Coupled Atmosphere Biogeochemical (CAB) model, which consists of a 1D global mean cloud-free, steady state atmospheric column model (with coupled climate and photochemistry), which calculates the temperature-, $H_2O$ vapor- and chemical species profiles, and a biogeochemistry submodule, which calculates aqueous $O_2$ concentrations. In this work, only the atmospheric column submodule was used since such biogeochemical processes on Earth-like planets are unconstrained. Furthermore, the resulting temperature and species concentration profiles were used by GARLIC (Generic Atmospheric Radiation Line-by-line Infrared-microwave Code) to calculate synthetic transmission spectra for the various hypothetical Earth-like planets studied.

Different "families" of scenarios have been used in this work. These include the modern Earth around the Sun, the AD Leo control run (with Earth's biomass but placed in the HZ around the M-dwarf star AD Leo), and the reduced $O_2$ and "dead" Earth scenarios (simulated by removing biomass). For the last two "families", i.e. the reduced $O_2$ and "dead" Earths scenarios, the $CO_2$ content as well as their sRHs have been varied, due to the ill-constrained nature of these parameters. Lastly, an $O_2$ parameter study has been performed for the hypothetical Earth-like planet with Earth's biomass around AD Leo, in order to determine the smallest atmospheric $O_2$ concentration necessary for the $O_2$ features to become visible in the planet's transmission spectrum.

In terms of atmospheric responses, this work has shown that interesting altitude-dependent chemical climate couplings appear in the $O_2$ concentration profile on separating the effects of lowering $O_2$ alone compared with lowering both $O_2$ as well as biomass emissions. The results suggest the largest differences in the lower layers but only small differences in the upper layers. This indicates that if life were not present, it could be hard to distinguish such worlds from planets with biospheres and low oxygen contents due to the abiotic production of $O_2$, posing a possible "false-positive" detection when searching for life. This statement assumes that $O_2$ detection is feasible. Our results however (see section 4.3) suggest that such a detection could be challenging over a range of $O_2$ abundances for e.g. the ELT and would likely be a task for future missions such as LIFE (Large Interferometer For Exoplanets) (Defrère et al., 2018).

Moreover, this work has shown that unexpected responses of $CH_4$ can occur even when separating the effects of lowering $O_2$ and lowering both $O_2$ and biomass contents due to feedbacks with its sink OH. In terms of the $H_2O$ concentration profile, changing the relative humidity on the surface of hypothetical Earth-like planets causes a visible difference in $H_2O$ vmr in the upper layers of the atmosphere compared to the AD Leo control run. On the other hand, the effect of changing the biomass is not as apparent, with the exception of the enhanced $CO_2$ scenarios where the high $CO_2$ content prevents $H_2O$ from being photolyzed as strongly.

Furthermore, the modern Earth around the Sun run exhibits smaller $N_2O$ vmr due to this molecule being photolyzed more efficiently by its stronger UV environment compared to the AD Leo control scenario. Despite removing the biomass at the surface for the "dead" Earth scenarios, the atmosphere still produces considerable $N_2O$ abiotically in the upper layers of the atmosphere so that its vmr is similar to that of the reduced $O_2$ scenarios in which biomass is present, leading to a possible "false-positive" detection by remote observations. Our results suggested that this behavior is shown to a smaller degree in the upper atmospheric regions for the enhanced $CO_2$ scenarios, making $N_2O$ for these cases in this sense a good biosignature.

Our work showed that $CH_3Cl$ displays a decrease in vmr with decreasing surface emissions and that changing the $CO_2$ content and surface relative humidity does not influence the $CH_3Cl$ vmr as much as removing the biomass. The stronger UVC environment for the modern Earth around the Sun scenario leads to reduced $CH_3Cl$ vmr for this scenario compared to the AD Leo control scenario. $CH_3Cl$ can be considered a good biosignature in the sense that there is a clear separation between the "dead" Earth and reduced $O_2$ scenarios in the upper layers of the atmosphere, where measurements are taken. Our work also showed that changing the sRH and $CO_2$ contents of a hypothetical planet around the M-dwarf star AD Leo influences the temperature-pressure profile more strongly than changing the $O_2$ and biomass contents.

In terms of spectral responses, our work has shown that changing the biomass alone produces a difference in effective height of up to 6 km, whereas changing the $CO_2$ content and sRH produces differences of up to 25 km in effective height. Thus, the latter produces larger differences in the transmission spectrum of a hypothetical Earth-like planet in the HZ of the M-dwarf star AD Leo. Furthermore, the $O_2$ parameter study has shown that in transmission spectra the $O_2$ feature is only apparent at 0.77 μm for $O_2$ concentrations of 0.1 PAL $O_2$ (corresponding to a value during the NOE about 800 million years ago) and 1 PAL (present value). A photosynthetic oxygen-producing biosphere which evolved on Earth prior to the GOE at about 2.7 Gyr on other Earth-like planets would be barely detectable for a very long period in time and only complex photosynthetic oxygen-producing life which evolved on Earth after the NOE would be visible in a transmission spectrum. The $O_3$ features in transmission spectra are clearly visible for $O_2$ concentrations of $10^{-2}$ PAL and higher which could indicate a photosynthetic oxygen-producing biosphere after the GOE. Our work has also shown the importance of considering biosignature responses in the region where they will likely by measured and the consequences for biosignature attributability.

Regarding detectability, results suggest that $CH_4$ is detectable with the ELT on Earth-like planets up to several tens of pc from the Earth except for our $10^{-4}$ and $10^{-6}$ $O_2$ PAL cases where strong UV led to $CH_4$ loss in the upper atmosphere. $O_2$ is challenging to detect even for the 1 PAL $O_2$ case and unfeasible for 0.1 PAL $O_2$ and less. It would be interesting to revisit these calculations assuming a transiting planet around Proxima Centauri or Barnard's star.

**Acknowledgements** John Lee Grenfell gratefully acknowledges ISSI Team 464 for useful discussion. This research was supported by DFG projects RA-714/7-1 and SCHR 1125/3-1.

**Author Disclosure Statement** No competing financial interests exist.